\documentclass[10pt, journal,compsoc]{IEEEtran}
\usepackage[T1]{fontenc}
\usepackage{xcolor}
\usepackage{float}
\usepackage{cite}  
\usepackage{enumerate}
\usepackage{hyperref}
\ifCLASSINFOpdf
   \usepackage[pdftex]{graphicx}
\else
\fi
\begin{document}
%
\title{A Survey on Exploratory Spatiotemporal Visual Analytics Approaches for Climate Science}
%
%
%
%

\author{
Abdullah-Al-Raihan Nayeem$^\dag$,
Dongyun Han$^\dag$,
Huikyo Lee,
Donghoon Kim,\\
Daniel Feldman,
William J. Tolone,
Daniel Crichton,
and Isaac Cho, \textit{Member, IEEE}
\IEEEcompsocitemizethanks{
\IEEEcompsocthanksitem AAR. Nayeem and W.J. Tolone are with the College of Computing and Informatics, University of North Carolina at Charlotte, Charlotte, NC 28223, United States. \protect
E-mail: \{anayeem, wjtolone\}@uncc.edu

\IEEEcompsocthanksitem D. Han, D. Kim, and I. Cho are with the Department of Computer Science, Utah State University, Logan, UT 84322, United States. \protect
Email: \{dongyun.han, donghoon.kim, isaac.cho\}@usu.edu

\IEEEcompsocthanksitem H. Lee, D. Crichton are with Jet Propulsion Laboratory, California Institute of Technology, Pasadena, CA 91109, United States. \protect
Email: \{huikyo.lee, daniel.j.crichton\}@jpl.nasa.gov

\IEEEcompsocthanksitem D. Feldman is with the Earth and Environmental Sciences Area, Lawrence Berkeley National Laboratory, Berkeley, CA 94720, United States. \protect
Email: drfeldman@lbl.gov
\IEEEcompsocthanksitem $^\dag$ These authors contributed equally
}

\thanks{Manuscript received April 19, 2005; revised August 26, 2015.}}

%
%

\markboth{Journal of \LaTeX\ Class Files,~Vol.~14, No.~8, August~2015}%
{Shell \MakeLowercase{\textit{et al.}}: Bare Demo of IEEEtran.cls for Computer Society Journals}
%



\IEEEtitleabstractindextext{%
\begin{abstract}
Climate science produces a wealth of complex, high-dimensional, multivariate data from observations and numerical models. These data are critical for understanding human-induced climate change and its socioeconomic impacts, as well as for effective decision-making for potential adaptations. Climate scientists are continuously evaluating output from numerical models against observations. This model evaluation process provides useful guidance to improve the numerical models and subsequent climate projections. 
Exploratory visual analytics systems possess the potential to significantly reduce the burden on scientists for traditional spatiotemporal analyses. In addition, technology and infrastructure advancements are further facilitating broader access to climate data. Climate scientists today can access climate data in distributed analytic environments and render exploratory visualizations for analyses. Efforts are ongoing to optimize the computational efficiency of spatiotemporal analyses to enable efficient exploration of massive data. These advances present further opportunities for the visualization community to innovate over the full landscape of challenges and requirements raised by scientists. 
In this report, we provide a comprehensive review of the challenges, requirements, and current approaches for exploratory spatiotemporal visual analytics solutions for climate data. We categorize the visual analytic techniques, systems, and tools presented in the relevant literature based on task requirements, data sources, statistical techniques, interaction methods, visualization techniques, performance evaluation methods, and application domains. Moreover, our analytic review identifies trends, limitations, and key challenges in visual analysis. 
This report will advance future research activities in climate visualizations and enables the end-users of climate data to identify effective climate change mitigation strategies.
\end{abstract}

}

\maketitle

\IEEEdisplaynontitleabstractindextext

%
\IEEEpeerreviewmaketitle


%
%
%
%

\IEEEraisesectionheading{\section{Introduction}\label{sec:introduction}}
\IEEEPARstart{I}{n}
recent years, climate science and visualization researchers have made significant advances in the spatiotemporal visual analysis of climate data. Interactive visualizations, today, enable scientists to hypothesize, investigate, and compare spatiotemporal fields from various observations and numerical models in innovative ways while leading to deeper, more insightful understandings. Nevertheless, the breadth of climate science and its diversity with numerous branches and sub-disciplines \cite{von1988theory, board2007earth} continuously demand visual analytics (VA) capabilities that can flexibly meet diverse sets of requirements \cite{borradaile2003statistics, Andrienko2003}. Climate data  are mostly spatiotemporal,  high-dimensional, and complex. As such,  to create cognitively efficient VA solutions for climate data, researchers often propose use-case specific approaches for visualizations and user interactions \cite{alam2021survey} rather than building versatile approaches to explore 
data.  

There are many spatiotemporal VA systems customized for various exploratory and predictive analyses in climate science. For example, several VA systems facilitate analyses of long-term climatology \cite{McLean2020, Alder2015, li2020sovas, Lichenhui2021} and focus on analyzing weather data \cite{wang2014, Sharma2018, wang2019open}. In addition, some systems feature specific aspects of climate data analysis such as uncertainty quantification \cite{Sanyal2010, MacIejewski2010}, anomaly detection \cite{Cao2018}, and adaptation and mitigation of climate change \cite{Herring2017, Johansson2017}. Moreover, spatiotemporal visualizations are often leveraged to highlight the temporal evolution of a data field by visualizing data trajectories. \cite{10.1093/llc/fqy084, PETERSON2020100316, 8440040}. General purpose visualization systems encapsulating data formats and analysis pipelines \cite {santos2013uv, 2018AGUFMIN21B35A, Huntington2017} for spatiotemporal data are also receiving attention in the earth science research community.

The aforementioned research efforts illuminate the challenges and requirements when developing spatiotemporal VA systems for climate data. The requirements commonly include designing a pipeline that supports data cleaning, data integration, data transformation, and dimensional reduction of data \cite{keimvascopes}. In addition, data are often resampled into a structured grid to facilitate visual presentation \cite{Fuchs2009}. All of the challenges and requirements are framed by various types of analysis tasks \cite{Keim2008}. Furthermore, interactive features are introduced in VA systems to facilitate filtering, exploration, and hypothesis generation of climate data \cite{Kehrer2008}. The complexity of this landscape and the diversity of research fields in climate science motivate the comprehensive review of spatiotemporal VA systems.

To date, a comprehensive review of the challenges, requirements, and current approaches to exploratory spatiotemporal VA solutions for climate science is missing. This report, however, presents a comprehensive survey and organizes relevant domain knowledge and VA solutions for climate science and visualization researchers. This report also outlines a VA pipeline adapted to exploratory analysis tasks to render interactive visualization over distributed spatiotemporal climate data. To promote innovation within the visualization community, this report serves as a guide to visualization experts by identifying challenges, requirements, trends, limitations, and gaps in existing spatiotemporal VA systems for climate science. At the same time, this report advances climate science by informing scientists and decision-makers of relevant spatiotemporal VA solutions, including their applicability and limitations, that could be used in support of policymaking and the identification of 
effective climate change mitigation strategies.

To achieve these goals, we reviewed current literature to summarize recent research advances in exploratory spatiotemporal VA solutions for climate science. Based on the literature review, we identified and characterized exploratory spatiotemporal VA systems along the following key dimensions \cite{keim2002, Kehrer2008}: 

    \textbf{Data Source \& Definition} – Climate data are commonly high-dimensional and often generated as high-resolution geospatial projections \cite{meehl2013climate, schmidt2014configuration}. We organize the data source \& definition dimension of the exploratory spatiotemporal VA domain according to the data structures and formats utilized by climate scientists \cite{edward2008experience, Davis2017}.
    
    \textbf{Statistical Analysis} – Climate scientists apply novel statistical analysis techniques to explore the potential impacts of climate change. The statistical analysis of variables such as precipitation, temperature, and those associated with air quality \cite{Lee2019, atmos12101239} is often accompanied by visualization. Along this dimension of the exploratory spatiotemporal VA domain, we organize solutions based on the method by which statistical analyses 
    are incorporated into the solution.
    
    \textbf{Visualization Techniques} – Novel exploratory visualization approaches \cite{ kappe2019analysis, Steed2020, Zahan2021} have been leveraged to create spatiotemporal representations for climate model evaluations relative to climate data observations and corresponding analysis tasks. Throughout this paper, we highlight the visualization techniques employed by current exploratory spatiotemporal VA for climate science. In addition, we assess visualization techniques for other spatiotemporal data \cite{10.1145/2093973.2094060, 8440040, 8613864} that may address requirements from underexplored areas of climate science.


    \textbf{Interaction Methods} – Interactive features facilitate scientists' efforts to explore, reconfigure, and compare the visual renderings of climate data. This dimension emphasizes the potential of each method to unfold knowledge from multivariate attributes (e.g., \cite{Fuchs2009, Tominski2021}).
    
    \textbf{Analytics Systems} – Novel analytics pipelines have been introduced and tested for VA systems (e.g. \cite{nayeem2021visual, dcpviz2022, KALO2020169}). We identify key features and limitations of the present pipelines in supporting climate data analysis.
    
    \textbf{Performance Evaluation} – It is challenging to define quantitative and qualitative metrics to measure the performance of VA systems \cite{Tominski2011cr, nocke2007visual}. Nevertheless, we organize this dimension of our review according to the respective categorization techniques, data types, and analysis tasks \cite{gerst2020using, mayr2018} to close the research gap between usability and task efficiency in climate science VA.

This paper is organized as follows. 
Section \ref{sec:previous-surveys} discusses the previous surveys related to visualizing climate data and spatiotemporal VA.
Section \ref{sec:methodology} presents our survey methodology, including how we collected and filtered peer-reviewed publications relevant to our topic.
Section \ref{sec:task-requirements} summarizes the landscape of task requirements for exploratory spatiotemporal VA solutions.
Section \ref{sec:data-sources} presents a summary of spatiotemporal data and exploratory analysis tasks in climate science.
Section \ref{sec:visualization-techniques} discusses the visualization techniques and interaction methods employed for spatiotemporal datasets.
Section \ref{sec:va-systems} examines the interface designs, tools, and infrastructures leveraged in existing systems and a pipeline for state-of-the-art VA approaches.
Section \ref{sec:evaluation-methods} discusses the qualitative and quantitative metrics to measure the performance of VA systems and close the research gaps between usability and task efficiency 
Section \ref{sec:discussion} overviews current trends in VA research related to climate science  and reflects on existing research gaps and potential future research directions to address these gaps and the ongoing challenges of understanding spatiotemporal climate data.

\section{Previous Surveys}\label{sec:previous-surveys}
We identify nine previous surveys that appear relevant to this report. In the following, we summarize these studies and highlight their significance. We also identify how this report advances the work of these previous studies.

Andrienko et al. \cite{Andrienko2003} reported an analytical review on exploratory spatiotemporal visualization. This work focuses on two important capabilities of visualization techniques - 1) exploratory analysis and 2) compatible data types. To relate the visualization tools and techniques to specific tasks and data, this report provides a categorization of exploratory analysis tasks based on 3W: where (space), when (time), and what (event) \cite{peuquet1994s}. Moreover, the user interaction techniques and the concept of a reconfigurable interface are discussed to support visual querying. The visualization techniques reported for long-term trend analysis are generally applicable to long-term climate data.

Tominski et al. \cite{Tominski2011cr} reported a survey interviewing 76 participants working in multiple fields of climate science. The authors collected information about the usage of visualization tools and analytical tasks of the participants 
to outline how an interactive visualization tool can benefit climate science research. They also inquired about the participants and discussed the reasons for the low pervasion of the state-of-the-art visualization tools to perform analysis tasks.

Rautenhaus et al. \cite{Rautenhaus2018} surveyed visualization tools and techniques for meteorological data analysis tasks. Their work highlights the categorization and summary of visualization approaches for operational forecasting with large-scale meteorological data. In this survey, a map visualization is identified as vital in visualizing both observational and numerical model data. This paper approaches the existing techniques and tools from two domain perspectives, visualization and meteorological, to assess the support for studying meteorological data.

Yang et al. \cite{yang2019research} classified visualization tools and analytical methods. Their paper focuses on enhancing human perception and improving cognition efficiency for spatiotemporal data analysis. It characterizes visualization as a series of conversion processes (raw data to visual artifacts) that recognize the human eye’s perception and the brain’s cognition abilities to gain knowledge visually. In this paper, spatiotemporal visualization methods are further categorized using statistical charts combined with map-based visualization methods, methods that are further classified according to five approaches - 1) Aggregation Charts; 2) Hierarchical Color Maps; 3) Heatmaps; 4) Origin-Destination Maps; and, 5) 3D Virtual Visualization.

He et al. \cite{he2019variable} illustrated trajectory analysis to extract patterns from multivariate spatiotemporal data. This paper conceptualizes the visualization of spatiotemporal trajectory on both geographic and abstract domains. They reported five encoding techniques (coordinated multiple views, juxtaposing, superimposing, cutting, and explicit encoding) to support visual comparison of trajectory data along with their corresponding weakness and strength. The interaction techniques are categorized based on the target objects such as view manipulation, data manipulation, transition, etc. rather than the interaction types.

Chen et al. \cite{CHEN2019129} reported the state-of-the-art multivariate VA techniques for simulation data. The visualization techniques are characterized into three categories. The first category is the visual designs in simulation space that covers the visual mapping and summarizing of the spatiotemporal data. The second category is the parameter space analysis for reasoning the model simulation that covers the visual steering, parameter space projection, etc. The third category is the data processing in feature space that covers the spatiotemporal feature definition, extraction, clustering, etc.

Rudenko et al. \cite{su14095248} conducted a brief review of the 4D visualization tools specialized for weather data. Their review summarizes the tools and techniques that support multi-format data from the processing and visualization phenomenon. However, their discussion reflected solely on the present advancements in geovisual environments while there are existing knowledge gaps of climate scientists' needs for exploratory analytics.

Andrienko et al. \cite{andrienko2020spatio} reported state-of-the-art spatiotemporal VA that summarizes the advancements of VA concerning spatiotemporal analysis. Their review focuses on the research efforts behind spatiotemporal visualization of trajectory movement, data filtering, aggregation, and spatial event clustering and distribution. A time-based data querying and filtering technique `Time Mask' is illustrated where temporal data are displayed in a space-time cube visualization. Moreover, this paper reflects on the shortcomings of the VA approaches in dealing with robustness and limited human cognition factor instead of overly focusing  on large data volumes.

Alam et al. \cite{alam2021survey} conducted a survey on spatiotemporal data analytics systems. They categorized their contributions into three groups - 1) spatial databases, 2) spatiotemporal data processing infrastructures, and 3) tools to support spatiotemporal analysis. The scope of this survey does not include specific visualization techniques to perform spatiotemporal analysis tasks. Rather, this work illustrates the spatiotemporal data management approaches, available processing infrastructures, and libraries/packages for VA.

The spatiotemporal VA approaches discussed in those surveys are applicable to climate science research, thus, have some overlaps with our paper. However, except for \cite{Tominski2011cr, Rautenhaus2018}, the other surveys do not scope any specific analytical use cases from the perspective of climate scientists. In contrast, our paper surveys the current progress of spatiotemporal visualization techniques and analytics approaches from the perspective of climate scientists. By incorporating this perspective, our survey makes two important contributions. First, this report advances climate science by informing scientists and decision-makers in a manner that aligns with use cases of relevant spatiotemporal VA solutions, including their applicability and limitations, that may be used in support of policymaking and the identification of appropriate and effective climate change mitigation strategies. Second, this report serves as a guide to visualization researchers by identifying challenges, requirements, trends, limitations, and gaps in existing spatiotemporal VA systems in the context of their use in climate science.


\section{Methodology}\label{sec:methodology}

This section presents our survey methodology. To conduct a survey of the state-of-art spatiotemporal VA in climate science, we reviewed literature from geographical science, climate science, visualization, and other related application domains. In doing so, we utilized a rich set of search terms, including the synonyms and variants of our topic. The search terms used included but were not limited to `climate', `spatiotemporal', `visualization', `exploration', and `analytics'. VA researchers on our team searched spatiotemporal visualizations-related literature from reputed peer-reviewed visualization and visual analytic publication venues. Similarly, climate scientists on our team searched for relevant climate science analytics literature from reputed peer-reviewed climate science journals and conferences. In addition, we jointly identified relevant literature based on spatiotemporal analysis published in reputed visualization venues (e.g., \cite{santos2013uv, Lichenhui2021, KALO2020169}). This approach provided us with a rich set of literature focused on spatiotemporal visualization and analytics in climate science. 

To collect the papers, we used the search term method previously described in different combinations using IEEE Xplore and ACM digital libraries. Our initial searches focused on papers published during the last decade. After constructing an initial collection, we began reviewing the titles and abstracts to filter the papers specific to our topic. Next, we reviewed the previously identified surveys and state-of-the-art reports (see section 2). These efforts expanded our pool of related research according to data types and climate science use cases. Then, we widened our search scope further to include additional relevant literature from the journals and conferences that previously had not been included in the initial list. At each stage, filtering was conducted as described above. 


After thoroughly filtering the collected papers, our resulting collection of relevant literature included 108 papers for further review. 
Once we reviewed these papers, we organized the papers using the taxonomy presented in section \ref{sec:task-requirements}. Next, we extracted the visualization techniques, analysis tasks, user interactions, and evaluation methods to organize the collection further. In addition, the climate scientists on our team identified, reviewed, and validated the climate science task requirements while the visualization experts on our team performed a similar task for the identified visualization requirements for exploratory spatiotemporal analysis. Research gaps and potential future research directions 
are presented in section \ref{sec:discussion}.



\section{Survey of Task and Visualization Requirements}\label{sec:task-requirements}
In addition to reviewing the recent advances in spatiotemporal visualization and VA approaches in climate science, we have studied analytical reviews and previously conducted domain experts' surveys related to our topic. Climate scientists advance the scientific understanding of how the Earth system achieves a climate state, and how that state has changed, is changing, or will change in the future, and consequently analyze large volumes of data, both produced by models and collected from observations. These datasets can be very large, can be homogeneous or heterogeneous, and can contain spatial and temporal dimensions, so visualization requirements for using these datasets to support and augment climate science must first recognize the specific data science challenges that climate science data present. These requirements must also recognize fundamental features that are expressed across observational and modeling climate datasets and also recognize that these datasets can expand at geometric rates.  

Additionally,  massive  climate data are often fragmented across distributed data servers \cite{nayeem2021visual}. Alam et al. \cite{alam2021survey} addressed the challenges of extracting and analyzing knowledge from the spatiotemporal data due to the heterogeneity and implicit spatial and temporal dependencies. It is widely popular among climate scientists to leverage high-performance clusters or supercomputers to enable exploratory analyses \cite{afzal2019state}.  Therefore, VA systems should address these storage and computational requirements. Moreover, generalized and reproducible workflows are being developed in order to filter, aggregate, classify, and analyze the movement of the spatiotemporal variables \cite{andrienko2020spatio, guivarch2022using}. To explore both raw and processed climate data, various visualization tools and techniques are utilized such as time series charts, bar plots, scatter plots, map visualizations, etc. \cite{Tominski2011cr, yang2019research}. Nonetheless, several previous studies reported the difficulty in performing analysis tasks running back and forth between the data analytics and visualization tools (e.g., \cite{afzal2019state}). We recognize the need for an interactive VA interface, which can obtain an enormous potential to ease climate scientists in conducting exploratory tasks.





Based on our literature review and collaboration with the climate scientists in our team, we identified the following task requirements for a spatiotemporal VA system to support a wide range of workflows and investigations currently performed by climate scientists.  Massive climate datasets are typically not analyzed in an unguided manner, rather, there are very specific questions that scientists seek to explore in large datasets. The guided and targeted approach to scientific inquiries where visualization is an important, and generally necessary tool, enables the identification of more generalized findings regarding climate data. Because of this, there are interrelated data management, analysis feature, and VA requirements that a literature review and scientist's surveys have revealed.  Briefly, these are:

\textbf{Data Management Requirements}
\begin{enumerate}[\bfseries R1]
    \item To enable efficient large-scale data encoding and retrieval to support analyses and interactive visualizations.
    \item To provide access to disparate and fragmented data from distributed data hosts to support cross-relation and impact study.
    \item To provide a cloud environment and features for data extraction, pre-processing, and transformation to support data preparation and visual exploration.
    \item To enable, via standardized APIs, interactive query-based data accessibility and mapping across disparate, multivariate datasets.
\end{enumerate}

\textbf{Analysis Feature Requirements}
\begin{enumerate}[\bfseries R1]
    \setcounter{enumi}{4}
    \item To enable the identification of underlying trends, patterns, and anomalies in spatiotemporal data.
    \item To support multivariate analysis to explore associations among the variables related to climate change effects and impacts.
    \item To enable comparative analysis and aggregation based on spatiotemporal granularity, clustering, and ensemble analysis.
\end{enumerate}

\textbf{VA Requirements}
\begin{enumerate}[\bfseries R1]
    \setcounter{enumi}{7}
    \item To mediate user interaction with the data servers and high-performance computing clusters to run analytical workflows.
    \item To provide customized interactive visualization techniques and analytics interface to explore and analyze climate data.
    \item To enable exploratory analysis at various scopes and scales with rich user interactions to understand multivariate climate data at different levels of abstraction.
\end{enumerate}

If these requirements are achieved, then visualization approaches can be powerful tools for enabling scientific investigations by streamlining hypothesis generation and testing. Often, they are not, however, visualization approaches developed years to decades ago where computational capabilities were far inferior to the present, are still being used.  Supplanting vestigial solutions with modern, improved solutions can take time, or in some cases, does not happen. With established workflows, even on old or sub-optimal visualization solutions, there is a high barrier to entry for novel solutions, even if they are superior to those already being used.  These challenges may also explain the low pervasion of VA tools in climate science research reported in previous studies \cite{Tominski2011cr}. We discuss adoption challenges in the potential future research directions in our discussion.


\section{Spatiotemporal Data in Climate Science}\label{sec:data-sources}
In this section, we discuss climate science analytics from the aspect of spatiotemporal data management, analysis techniques, and applications. We present the definition of spatiotemporal data and its different transformations in the context of climate science. We scope the exploratory spatiotemporal analyses that are employed to study climate change, its impact, and possible adaptations. In turn, we introduce the analysis techniques and different applications of spatiotemporal climate variables.



\subsection{Data Definition}
\begin{figure}[t]
    \centering 
    \includegraphics[width=\linewidth]{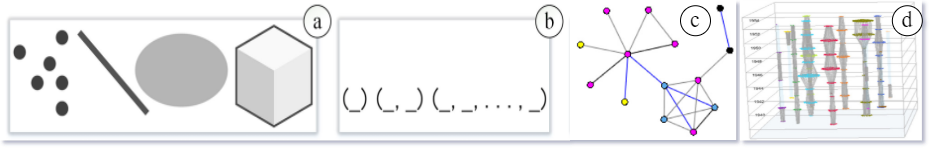}
    \caption{A conceptual illustration of different data types classified in \cite{Ben1996}. He et al. \cite{he2019variable} illustrated the concept of the data type classification: (a) Multidimensional: 0D, 1D, 2D, and 3D; (b) Multivariate: scalar, 2-tuple, and n-tuple. (c) Network and (d) Tree data types are illustrated from the view diagram presented by Tominski et al. \cite{Tominski2021}.}
    \label{fig:data-types}
\end{figure}

Events in spatiotemporal data consist of space and time. In simple terms, spatiotemporal data denotes events related to location that occurs over time. Climate science produces diverse spatiotemporal data based on the dimensions, events, and structure. We categorize different data types and sources employed in the VA interfaces according to the data type taxonomy discussed by Shneiderman \cite{Ben1996}. This taxonomy was later utilized by Afzal et al. \cite{afzal2019state} for the ocean and atmospheric datasets. The data type taxonomy proposed by Shneiderman includes 1-2-3 dimensional, temporal, multivariate, tree, and network data \cite{Ben1996}. He et al. \cite{he2019variable} presented a conceptual illustration of multidimensional and multivariate data among other types as shown in Fig. \ref{fig:data-types}a and b. In addition, we leveraged the data representation diagram presented by Tominski et al. \cite{Tominski2021} to illustrate the tree and network data types as shown in Fig. \ref{fig:data-types}c and d. 

\textbf{1-Dimensional (1D):}
This data type represents linear data such as textual documents, a list of names, and source codes. The data are organized in a sequential layout, generally consisting of unstructured information. As a result, it is particularly challenging to design visualizations for this data type. We did not find any instance of this type in context to spatiotemporal climate variables.

\textbf{2-Dimensional (2D):}
This data type represents planar data (i.e., x and y) that include a geographic map, floorplans, or layouts. Datasets of this type generally have additional attributes such as name, value, etc. 2D climate data often represents the spatial distribution of a variable \cite{santos2013uv, li2020sovas}. VA interfaces in climate science visualize these datasets with a multi-layer approach on a 2D map. Users can toggle layers of the map based on the variable of interest \cite{MacIejewski2010, Chang2016, Mohammadi2017}.

\textbf{3-Dimensional (3D):}
This data type includes 3D models of real-world objects (e.g., buildings, Computer-Aided Design (CAD) models) and scientific datasets (e.g., computed tomography scan). The data often contain attributes with volume and complex interrelation with other data items. Volumetric rendering is a popular technique to visualize scientific data \cite{helbig2014, Zhang2019, Li2019}.
Most modern climate models produce 3D data to study cross-correlation maps among climate variables \cite{Sukharev2009}. Climate change impact and adaption-related analysis also produce this type of data \cite{wang2018}.

\textbf{Multidimensional:}
Multidimensional data denotes multiple dimensions in the dataset whereas multivariate data denotes the multiple attributes associated with each dimension. Climate scientists study a wealth of multivariate data \cite{Tominski2021} to analyze variable association \cite{wang2018, anayeem2022}, distribution \cite{poco2014}, path ensemble \cite{li2020sovas}, etc. Multidimensional data is usually visualized using different 2D techniques such as integrated multiple visualization setup \cite{Huntington2017, Porter2021, anayeem2022}, small multiples \cite{poco2014, nayeem2021visual, dcpviz2022}, color/glyph stylization \cite{Hadlak2010, Zahan2021}, or summarized form with fewer dimension \cite{eaglin2017space}. Parallel coordinates are widely used for visualizing multidimensional and multivariate datasets \cite{wang2018, li2020sovas}.

\textbf{Temporal:}
This data type denotes datasets that contain time-varying variables. Temporal data denotes the changes in climate variables over time such as temperature, precipitation, humidity, etc. Data sources utilized in most of the VA interfaces discussed in this paper relate to this data type \cite{KALO2020169} as the temporal dimension contains an essential portion of spatiotemporal data. Temporal data is generally visualized using a standard 2D visualization technique such as area plot \cite{Lichenhui2021}, scatter plot \cite{Maskey2020}, animation \cite{Sharma2018}, time sliders \cite{Herring2017, Johansson2017}, etc. based on the data transformation.


\textbf{Tree:}
This data type represents hierarchically structured data where each item in the collection has a link to the parent item except for the root. The data items are often linked based on multiple attributes. These datasets are visualized using the different representations of tree/hierarchical visualizations (e.g., edge-bundling, treemap). Kappe et al. \cite{kappe2019analysis} detailed a hierarchical clustering based on the climate states and time-dependent ensembles (Fig. \ref{fig:cmv-technique}D). They used dendrogram and hierarchical clustering to visualize the data. Nayeem et al. \cite{nayeem2021visual} summarized spatiotemporal precipitation and temperature data to produce a hierarchical seasonal-regional mean. However, we understand that tree-structured data are not 
common in climate science as we found only a few examples. 

\textbf{Network:}
This data type includes the links among the data attributes that cannot be represented using a tree structure. Network data items often obtain an arbitrary number of relationships with other items that are commonly visualized using node-link diagrams and matrix representations. Kalo et al. \cite{KALO2020169} studied spatiotemporal interpolation of air pollution based on the data collected at measurement sites across the United States (US). They used a node-link diagram to visualize the triangulation of air pollution data (Fig. \ref{fig:trajectory-vis}A). Analysis of path ensemble \cite{liu2015visualizing}, spatiotemporal trajectory \cite{lusifei2011}, variable similarity \cite{poco2014, anayeem2022}, data flow \cite{kim2018} etc. derive network data for climate variables.


\subsection{Data Storage}
Storage of spatiotemporal data has been a challenge for the analyst and research community due to its volume and complex nature \cite{edward2008experience}. Climate data are characterized as complex because of high dimensionality, mutlivariability, and multi-resolution factors \cite{wang2018}. Advancement in infrastructure and equipment has made it possible to acquire a more granular level of spatiotemporal data, which demands vast and faster storage units. For faster transactions, spatiotemporal data are frequently stored in structured relational databases (e.g., PostgreSQL \cite{obe2017postgresql}, PostGIS \cite{obe2021postgis}). However, a large volume of data from multiple sources are often stored in a non-structured format, known as NoSQL (e.g., Redis \cite{da2015redis}, MongoDB \cite{chodorow2013mongodb}). Storage is still a concern for spatiotemporal data for efficient data encoding and retrieval.

In climate science, increasing resolution of geospatial data and frequent temporal records make Network Common Data Form (NetCDF) \cite{Rew1990} a well-accepted data storing format in the research community for multidimensional climate variables such as temperature, precipitation, and air quality \cite{edward2008experience, Eyring2016CMIP6, Scarponi2018}. NetCDF provides an encoding convention for the researchers to store multidimensional data in a relatively smaller space \cite{Davis2017}. Moreover, it allows streaming data without copying or defining the structure from scratch. Individual data files are simultaneously accessible by one writer and multiple readers \cite{Rew1990}. Climate scientists established metadata conventions for storing climate variables in NetCDF files \cite{eaton2003netcdf}. The conventions include acceptable data types, naming conventions, dimensions, units of measurement, etc. The metadata container should also describe the title, data source, institute, and necessary references for the data \cite{eaton2003netcdf}. Hence, NetCDF has become one of the widely used formats for storing earth science data \cite{Biard2016}.

Besides, based on general usage and data characteristics, spatiotemporal datasets are also stored as fragmented, partitioned clusters across distributed servers \cite{kim2018, nayeem2021visual}. As a result, exploratory analysis performed by researchers and scientists does not always reside in a single data server. Important insights and knowledge often reside in distributed and disparate datasets. Spatiotemporal data fragmented across multiple remote data hosts are identified as distributed spatiotemporal datasets. Big data computing infrastructure such as Hadoop \cite{5496972} supports large-scale data encoding
to enable distributed analysis \cite{Sharma2018}. Other big data infrastructures are being explored (e.g., Virtual Infrastructure and Virtual Information-Fabric Infrastructure (VIFI) \cite{8397642}), that mitigate analyses of distributed data.


\begin{table*}
     \centering
  \caption{Distribution of exploratory spatiotemporal analysis tasks supported by the current VA systems.}
  \includegraphics[width=0.95\linewidth]{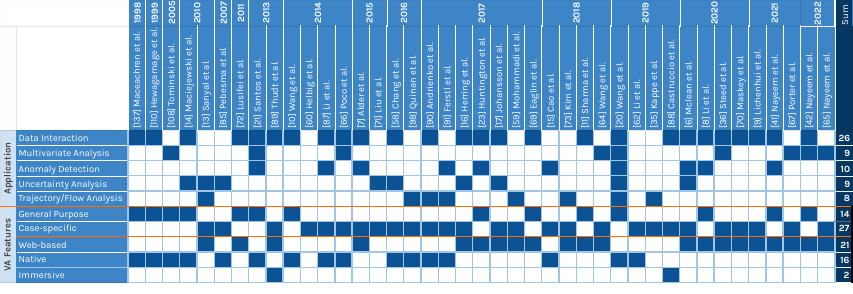}
  \label{tbl:va-applications}
  \vspace{-5mm}
\end{table*}

\subsection{Analysis Tasks}
A core element of future weather (or air quality) forecasts and climate projections are numerical models that not only provide a foretelling of physical indicators of future climate but also indirectly provide information on societal impacts and thus provide a key resource for addressing adaptation and mitigation questions. Because of the critical spatiotemporal data such models offer, it is a high priority to bring as much observational scrutiny to the output from the numerical models as possible. This requires the systematic application of observational datasets from various sources. As such, enabling spatiotemporal analysis of observational datasets and evaluation of the spatial-temporal output from numerical models are all necessary for visual analytic systems to provide a reliable characterization of future weather/air quality/climate that can lead to an informed decision-making process \cite{guivarch2022using}. 

\textbf{Data Exploration:} Uncovers valuable insights from the spatiotemporal data as one of the earlier stages in the analysis process to identify a subset/area of interest. Piltner et al. primarily focused on spatiotemporal interpolation of irregularly spaced air quality observations for the contiguous US \cite{KALO2020169}. Their spatiotemporal interpolation and its validation are conducted in the stage of data processing, so users of the visual analytic system explore the interpolated data via a web interface. Similarly, DDLVis \cite{Lichenhui2021} applies three advanced statistical techniques (a peak-based kernel density estimation, a dictionary learning method, and a peak-based variation generation model) to store, visualize, and query climate data efficiently. Climate Engine \cite{Huntington2017} provisions calculation of basic statistics including mean, median, maximum, minimum, and total only. 
Moreover, with special emphasis on the US National Climate Assessment (NCA), the National Climate Change Viewer (NCCV \cite{Alder2015}) provides various summary statistics for various geographical regions, such as counties, states, and NCA regions, over the contiguous US. The summary statistics include spatially averaged time series and percentile tables of temperature and precipitation from 30 models for present and future climate. 

\textbf{Multivariate Analysis:} In most analysis tasks, climate scientists analyze multiple variables which are associated with certain phenomena. For example, concentrations of ozone, nitrous oxides, and particulate matter from observations and numerical models are analyzed in air quality studies. In multivariate analyses of climate data, it is always important to figure out a normal state, which is usually defined as averages of individual variables over a certain period. Wang et al. \cite{wang2018} applied an association rule-learning algorithm to study the relationship of multiple variables in climate data. To facilitate the rule-learning algorithm and parallel coordinate plotting, they also applied a categorization algorithm to group each variable's values into five categories. Moreover, Poco et al. \cite{poco2014} demonstrated an inter-comparison among simulated climate models leveraging multifaceted data. CrossVis \cite{Steed2020} provides a VA system to explore large-scale heterogeneous multivariate data with a use case of historical hurricane observations.

\textbf{Anomaly Detection:} Anomalous events, such as heatwaves, cold surges, heavy precipitation, drought, and severe air pollution, can be detected by analyzing the deviation of spatiotemporal data from their averages. Using the Scalable Online Visual Analytic System (SOVAS; \cite{li2020sovas}), climate scientists can detect anomalous events such as extreme heat events. SOVAS also enables the calculation of correlation coefficients between two meteorological variables and spatial means. Voila \cite{Cao2018} presented a tensor-based analysis algorithm for interactively detecting anomalies in spatiotemporal data. Moreover, the phenomena portal \cite{Maskey2020} provides a map-based user interface to visualize certain anomalous events detected by their convolutional neural network (CNN) model. Users of the portal can provide feedback on the detected events.

\textbf{Uncertainty Analysis:} The evaluation of numerical models against observations aims to quantify uncertainty for future projections from the models. This uncertainty quantification process is key to informing climate model development and providing actionable climate information to support decision-making \cite{pebesma2007}. To visualize forecast output from multiple simulations and their uncertainty, Noodles \cite{Sanyal2010} applies two visualization techniques, uncertainty glyphs, and an uncertainty ribbon. The authors compared these two techniques with conventional spaghetti plots and show the advantages of the new techniques. Herring et al. \cite{Herring2017} present a context-switching technique in ClimateData.US to explore low and high-emission scenarios of climate variables such as temperature and precipitation. ClimateData.US essentially enables the uncertainty analysis of climate risks.

\textbf{Trajectory \& Flow Analysis:} Trajectory analysis are widely used to track hurricanes or air pollutants including volcanic ashes and dust storms. Kim et al. \cite{kim2018} developed a novel flow analysis technique to extract the flow map from the non-directional spatiotemporal data. EnsembleGraph \cite{Shu2016} utilizes graph visualization of ensemble simulations. In the graph, nodes are subregions with similar simulation output and edges represent spatial overlap. The Hawaii Rainfall Analysis and Mapping Application (HI-RAMA, \cite{McLean2020}) applies a random forest model for quality control of rainfall station data and two different weighting approaches for the gap-filling of missing data. Again, these statistical techniques are not applied to users' data exploration.





\begin{table*}
    \centering
    \caption{Categories for the visualization techniques reviewed in this paper. Standard 2D charts and geospatial contours have been popular choices in visual analysis. Coordinated Multi-Views are widely used in recent years to visualize spatiotemporal data.}
    \includegraphics[width=\linewidth]{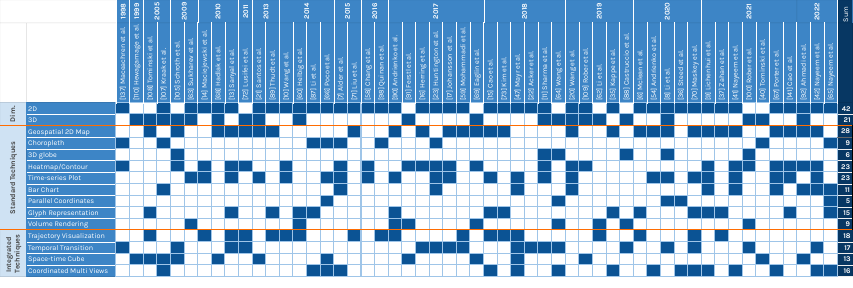}
    \label{tbl:cat-vis-techniques}
    \vspace{-5mm}
\end{table*}

\section{Visualization Techniques}\label{sec:visualization-techniques}
In this section, we discuss various visualization techniques that have been employed to visualize the spatiotemporal climate data. We investigated these techniques from different aspects to report a categorization and mapping to spatiotemporal data types and analysis tasks discussed in section \ref{sec:data-sources}. We reviewed the taxonomy for visualization techniques defined by Keim \cite{keim2002} and extended further by Afzal et al. \cite{afzal2019state}. In their work, the visualization techniques are classified as standard 2D/3D displays, geometrically transformed displays, icon-based displays, dense pixel displays, and stacked displays. However, climate visualizations, in many instances, leverage multiple visualization techniques to illustrate the spatiotemporal data due to its high dimensional and multivariate attributes. Hence, display design is also a crucial aspect of data visualization in climate science. In our review, we provide a classification of the visualization techniques inspired by \cite{keim2002}, albeit based on the display design instead of visual encoding. The classification focuses more on the different approaches of integrated visualization methods for 
spatiotemporal data as depicted in Table \ref{tbl:cat-vis-techniques}. 


To characterize the visualization techniques for spatiotemporal climate data, we first discuss the different standard 2D/3D visualizations such as time series plots \cite{Maskey2020}, bar charts, parallel-coordinates \cite{poco2014, Steed2020}, volumetric rendering \cite{wang2018, Li2019}, map visualizations \cite{Mohammadi2017, castruccio2019visualizing}, etc. Then, we discuss different display designs for integrated spatiotemporal visualizations for climate science such as temporal transition, path ensembles \cite{Papers2013, ANDRIENKO201725}, trajectory \cite{Li2019}, space-time cube \cite{Ferstl2017, Ahmadi2021}, radial map \cite{Li2014}, and coordinated multiple views \cite{nayeem2021visual, Lichenhui2021}. These integrated visual representations assist climate scientists in performing exploratory visual analysis such as sensemaking \cite{McLean2020}, anomaly detection \cite{Cao2018}, trajectory analysis \cite{kim2018}, and visual query \cite{Lichenhui2021}.

\subsection{Standard Visualization Techniques}
Climate variables generally contain timestamps and geolocation, so climate scientists widely use geospatial visualization and time series charts to analyze their data \cite{nocke2008visualization}. Geospatial visualization provides an interface to inspect and explore spatial variables with geographic location \cite{Cartwright2000, Alder2015, nayeem2021visual} whereas time series charts are essential to perceive the temporal trends of the spatial features. The most commonly used techniques to visualize spatiotemporal climate data are a combination of standard charts, such as a line chart, a scatter plot, and their variations, \cite{Rautenhaus2018, afzal2019state} along with a map visualization (i.e., satellite \cite{castruccio2019visualizing}, choropleth \cite{MacIejewski2010}). These standard visualizations are illustrated in 2D and 3D. The basic idea of 2D visualizations in this context is to present the data variable against time in x-y axes. In contrast, 3D visualization includes another dimension essentially supporting the analysis of trends and patterns in spatiotemporal data with x-y-z axes. However, leveraging standard charts often create occlusion on the visual interface that challenges cognition efficiency \cite{nocke2007visual}. In this section, we discuss different 2D and 3D visualization techniques for the visual analysis of spatiotemporal climate data.

\subsubsection{2D Techniques}


In climate science VA, 2D visualizations are the most commonly used technique to illustrate spatiotemporal data. This category includes basic charts such as line graphs, scatter plots, bar charts, etc. \cite{MacIejewski2010, Sanyal2010, Li2014, Huntington2017, Cao2018, Maskey2020}. Li et al. \cite{li2020sovas} used a time-series plot to present temporal trends and correlations between temperature and precipitation. Sharma et al. \cite{Sharma2018} leveraged a combination of a line graph and a scatter plot in their work to study the temperature anomalies in the ocean and land surfaces.

\begin{figure}
    \centering
    \includegraphics[width=1.0\linewidth]{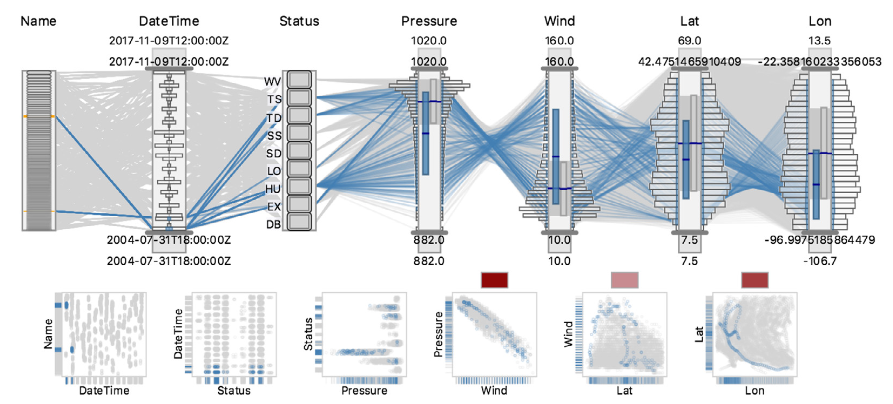}
    \caption{CrossVis demonstrates interactive axes selection of categorical cells and numerical range in parallel coordinates to explore large-scale multivariate data \cite{Steed2020}.}
    \label{fig:pcoords}
\end{figure}

Other 2D charts besides line and scatter plots are also utilized to visualize the spatiotemporal climate variables.  Parallel coordinates are popularly leveraged to analyze the association multivariate climate data (Task requirement \textbf{R6}). Poco et al. \cite{poco2014} studied the multi-model distribution of spatiotemporal climate variables in a parallel coordinates visualization. Steed et al. \cite{Steed2020} demonstrated interactive parallel coordinates consisting of categorical and numerical axes to explore large-scale heterogeneous multivariate data as shown in Fig. \ref{fig:pcoords}. Wang et al. \cite{wang2018} and Li et al. \cite{li2020sovas} also used a multifaced parallel coordinate plot to illustrate the association of climate variables near a cyclone center. Bar charts are also leveraged to compare the multi-model, seasonal, and yearly distribution of climate variables in \cite{Alder2015, Huntington2017, nayeem2021visual}. Various designs of statistical bar visualizations are depicted in Fig. \ref{fig:pcoords} and \ref{fig:cmv-technique}. In addition, Shu et al. \cite{Shu2016} introduced a 2D temporal graph that provides spatiotemporal behaviors in ensemble simulation data. Similarly, there are other 2D visualization techniques utilized for exploring spatiotemporal climate variables such as node-link diagram \cite{Tominski2021}, area plot \cite{pahins2020real}, and hierarchical clustering \cite{kappe2019analysis}.

The standard 2D charts are frequently presented as small multiples to allow the users to easily compare the climate variables against a timeline. Small multiples combine a set of basic visualizations (e.g., line graphs, bar charts) and encode data using the same scale to ease the visual comparison tasks \cite{tufte1990envisioning}. A specific representation from the basic charts and an arrangement method such as grid or radial need to be determined to implement small multiples \cite{ko2016survey}.  Small multiples are also leveraged to form a correlation matrix generally using scatter plots \cite{li2020sovas} or color-coded glyphs \cite{Tominski2021}. Examples of small multiples for comparing spatiotemporal climate variables are presented in \cite{poco2014, nayeem2021visual}.

In addition, we reviewed various implementations of 2D map visualization for a spatial portion of the climate data. Geospatial maps and choropleths are popularly leveraged to visualize the spatial intensity or projection of climate variables \cite{Quinan2016, Huntington2017, Maskey2020, Tominski2021}. Sliding animation and display transitions are generally implemented to support browsing the temporal data variables through the spatial visualizations \cite{Sanyal2010, Herring2017}. Zahan et al. \cite{Zahan2021} introduce 4 contour-line stylization techniques to study multivariate information such as temperature, elevation, etc. The encoding techniques include parallel lines, color blending, pie, and thickness-shade. Heatmap, contour, and glyph-based representations are also extensively used over geospatial maps as an additional layer to present the spatial intensity \cite{Johansson2017, McLean2020} (Fig. \ref{fig:trajectory-vis}C.)

Area maps are used in NASA's Giovanni and VAPOR to plot the area of modeled impact \cite{2007EOSTr8814A, Li2019}. Histogram and scatter plot have been leveraged in National Climate Change Viewer (NCCV) to compare the climate models by predicted max temperature \cite{Alder2015}
In the spatiotemporal VA interfaces for climate science, 2D standard charts are frequently implemented in an integrated fashion to encode high dimensional data and reduce occlusion \cite{Alder2015, Cao2018, Lichenhui2021}.

\subsubsection{3D Techniques}
\begin{figure}
    \centering
    \includegraphics[width=\linewidth]{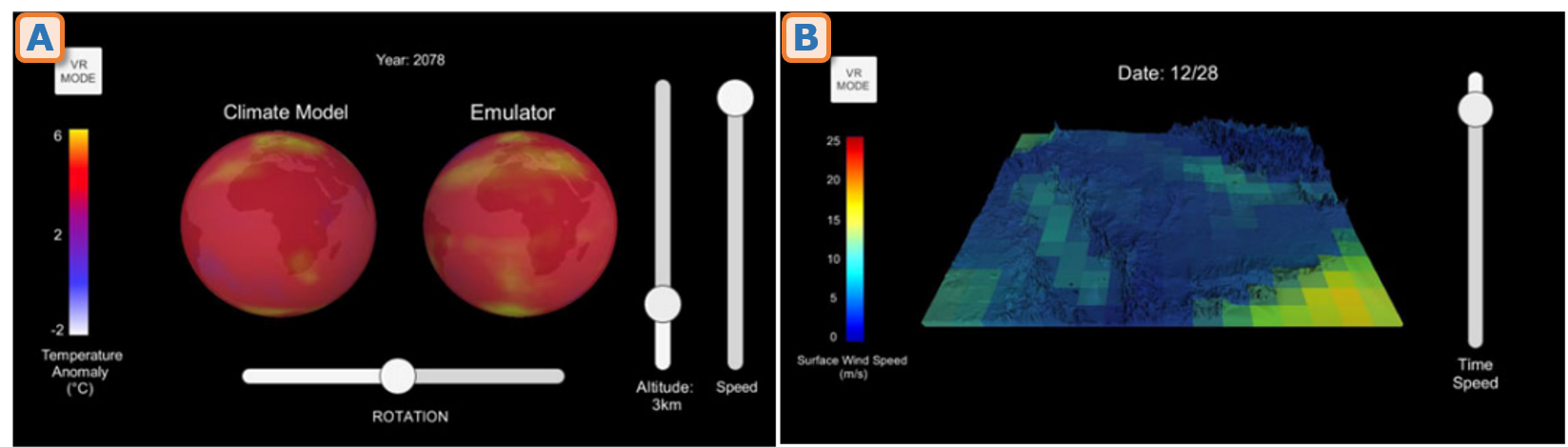}
    \caption{CORNEA \cite{castruccio2019visualizing} demonstrates an immersive virtual environment to explore simulated climate models using (a) globe and (b) surface display.}
    \label{fig:vr-vis}
\end{figure}

3D visualizations are also standard techniques demonstrated in many climate science VA interfaces \cite{Tominski2011cr, santos2013uv, Sharma2018}. These visualizations often cases provide more context than 2D techniques. Geospatial maps are widely used 3D visualizations for analyzing climate variables. Li et al. \cite{li2020sovas} utilized a 3D globe heatmap to present the worldwide heatwave. Rober et al. \cite{rober2021visualization}, in another work, discusses a 3D globe to demonstrate 21st-century projections of the climate variables. 
In addition, the use of variable raytracing and depth of field are demonstrated in \cite{rober2021visualization, castruccio2019visualizing} to direct the user's attention to a specific region on the 3D map (Fig. \ref{fig:vr-vis}b).


Moreover, volume rendering is another 3D visualization technique used in climate science VA \cite{Sukharev2009, Ma2009, wang2018}. Helbig et al. \cite{helbig2014} present a volume rendering to visualize the elevation, simulated precipitation, and mass fraction of snow. 3D glyphs are used in their encoding to illustrate the direction. Rober et al. \cite{rober2021visualization} discuss the visualization of 3D liquid cloud water and cloud ice using OSPRay volume rendering \cite{wald2017}. In addition, Zhang et al. \cite{Zhang2019} provide a rendering comparison to report an efficient dynamic volumetric encoding for meteorological data in a 3D globe. Li et al. \cite{Li2019} demonstrated an exploratory volumetric visualization in VAPOR, to interactively analyze the simulation data.  Wang et al. \cite{wang2019open} illustrated volume rendering over map visualization to study multidimensional meteorological data.

An immersive environment allows the user to interactively explore the spatiotemporal variables on a spherical shell. Castruccio et al. \cite{castruccio2019visualizing} presented a virtual reality environment, CORNEA, to visualize and compare the simulated climate models as shown in Fig. \ref{fig:vr-vis}. Helbig et al. \cite{helbig2014} also presented their volume rendering in a virtual reality environment using Autodesk VRED \cite{autodesk}.

\subsection{Integrated Visualization Techniques}
The display design for climate science VA is crucial as the complex spatiotemporal data often challenges human cognition. Several studies investigated the design factors that impact the human ability to comprehend the illustrated data \cite{nocke2004methods, Rautenhaus2018, mayr2018}. As a result of the multidimensional and multiresolution characteristics of the spatiotemporal data, VA environments often leverage multiple visual components integrated with interactive features to serve exploratory purposes. In this section, we discuss the integrated visualization techniques employed to explore the climate variables.

\subsubsection{Temporal Transition}
The transition technique in spatiotemporal visualization frequently consists of a geospatial or contour-based visualization to plot the spatial intensity along with a secondary visualization to stream through the temporal dimension. It provides a dynamic representation as the temporal events appear and disappear on the spatial map controlled by time or user selection \cite{mayr2018, Sharma2018}. The spatial dimension illustrated in 2D/3D geospatial maps is often layered with a base map or choropleth \cite{Herring2017, dcpviz2022, anayeem2022}. The temporal progression of data variables is summarized using standard visualization techniques such as line graphs, area charts, scatter plots, or a series of geospatial depictions of the temporal data \cite{pahins2020real, Maskey2020, nayeem2021visual, Lichenhui2021}. Time sliders are also a popular choice to demonstrate the temporal transition on a geospatial visualization for spatiotemporal data \cite{schroth2009tools, Sanyal2010, schmidt2014configuration, Herring2017, li2020sovas}. Display transition and animation have been employed in CORNEA \cite{castruccio2019visualizing} where sliders are implemented to control the time speed, rotation, and altitude on the 3D globe. This technique is particularly useful to explore smaller slices from large-scale spatiotemporal data. The user can interfere in the transition at a time and select a location to observe the changes/characteristics of the particular variable (Fig. \ref{fig:vr-vis}.)

Maskey et al. \cite{Maskey2020} demonstrated a spatiotemporal depiction using the display transition of geospatial visualization. Pahins et al. \cite{pahins2020real} designed an integrated visualization to explore large-scale spatiotemporal data using a geospatial view and area chart. The transition view allows the user to perceive the subtle temporal change of events in space. However, the fast progression of events can overwhelm the user when exploring the data variables. User attention is also split between the geospatial and timeline views which might lead to missing the non-salient changes over time. Mayr et al. \cite{mayr2018} discussed several studies to compare the animation speed and user interactions for spatiotemporal transition to find cognitive efficiency. 

Temporal transitions are also utilized in flow visualizations. Lu et al. \cite{lusifei2011} demonstrated the use of temporal transition to visualize the flow of sea level pressure. Liu et al. \cite{liu2015visualizing} visualized the hourly storm path ensembles transitioning the display. Li et al. \cite{Li2019} illustrate the animation of volume rendering to present a simulated progression of the hurricane path.

The transition technique is usually mapped to the time dimension to depict the spatial change of events over time. This technique fosters human cognition to detect subtle temporal changes. However, a fast transition on a large spatial dimension can deprive the user to miss non-salient changes. User-preferred transition speed and time step selection can resolve this issue.

\subsubsection{Space-time Cube}
\begin{figure}[t]
    \centering
    \includegraphics[width=\linewidth]{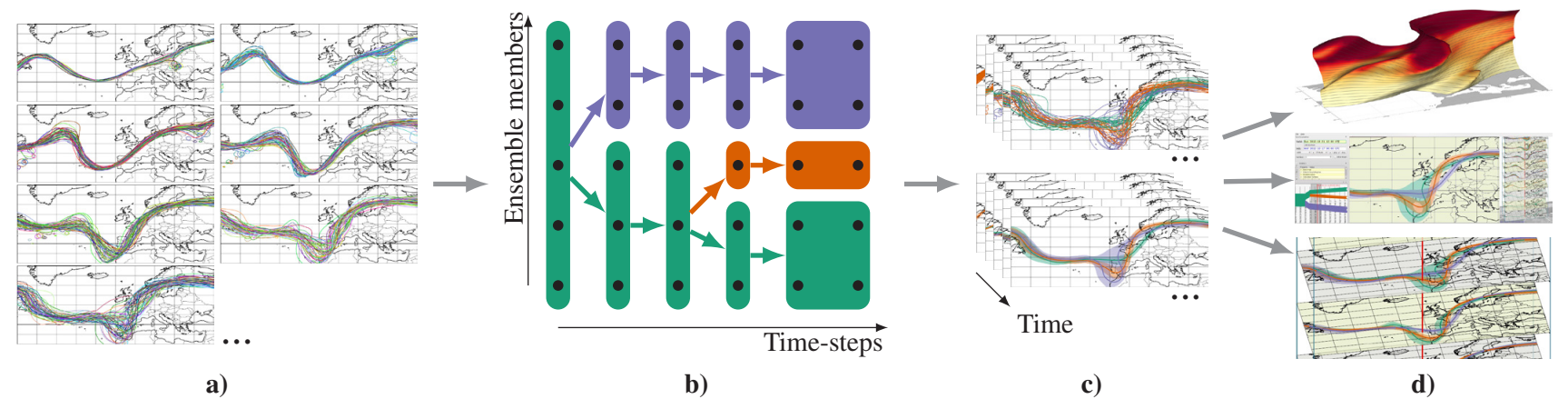}
    \caption{Ensemble of time-varying iso-contours in a weather forecast to produce interactive space-time volume rendering through time-hierarchical clustering \cite{Ferstl2017}.}
    \label{fig:space-time-cube-vis}
\end{figure}

Space-time cube is an integrated 3D visualization technique leveraged in visualizing various spatiotemporal climate variables. To answer the question of where (space) and when (time) from the spatiotemporal climate data, Kraak \cite{kraak2003space} presented this technique combining specific visualizations for space and time. In a space-time cube, the cube's bottom represents space whereas the height represents time with temporal cutting operations. A spatial map is generally used to denote the space followed by multiple layers of standard visualizations to illustrate the climate variables in different timestamps in the dataset \cite{kraak2005timelines, Tominski2005, andrienko2020spatio}. We review the use of bar plots, scatter plots, volume rendering, and spatial maps on the temporal axis in space-time cube visualizations. 

\begin{figure}
    \centering
    \includegraphics[width=0.8\linewidth]{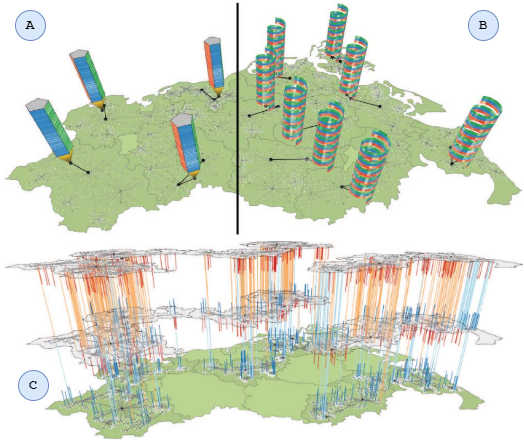}
    \caption{Visual hierarchy of multivariate spatiotemporal data demonstrated by Hadlak et al. \cite{Hadlak2010} using glyph-based space-time illustration.}
    \label{fig:spiral-vis}
\end{figure}

Schroth et al. \cite{schroth2009tools} conducted a case study on the existing tools for understanding the spatiotemporal climate scenarios where they presented a space-time cube for resolving the multi-dimensional climate data visualization problem. Eaglin et al. \cite{eaglin2017space} provided a web-based interface illustrating a space-time cube where the temporal axis consists of a volume rendering. The user can interact with the 3D view to visualize a time layer from the volume in a 2D heatmap. Poco et al. \cite{poco2014} demonstrated the use of a space-time cube in the SimilarityExplorer to visualize the complexity of multifaceted climate data variables. Hadlak et al. \cite{Hadlak2010} presented various illustrations of space-time cubes using colored links (Fig. \ref{fig:spiral-vis}), pencil, and helix glyphs to visualize temporal patterns in spatiotemporal variables. Visual encodings for the temporal axis are able to include multivariate climate data in a space-time cube visualization. Rober et al. \cite{rober2019} utilized a space-time cube representation while demonstrating in-situ visualization \cite{Ma2009} processing with climate data. Ferstl et al. \cite{Ferstl2017} presented an approach to visualize time-varying iso-contours in a time-hierarchical clustering using juxtaposition as shown in Fig. \ref{fig:space-time-cube-vis}. The stacked iso-contours in the space-time cluster essentially form a volumetric rendering of the weather forecast ensembles. 

Hewagamage et al. \cite{hewagamage1999interactive} illustrates a spiral view on the geospatial map to visualize the spatiotemporal pattern. This concept was later extended by Tominski et al. \cite{Tominski2005} where they capitalized the spiral helix to visualize multivariate spatiotemporal data to explore trends and correlations among attributes as shown in Fig. \ref{fig:spiral-vis}.

Space-time cube visualizations are able to encode high-dimensional spatiotemporal data in an integrated display design. Hence, this visualization is utilized to provide an overview of the spatiotemporal events to the users. However, with an increasing number of time steps and variables in the data, the visual occlusion can challenge user cognition when conducting the exploratory analysis. Select interaction aligned with the temporal axis is discussed as one of the potential solutions in \cite{eaglin2017space, andrienko2020spatio}.

\begin{figure}[t]
    \centering
    \includegraphics[width=1.0\linewidth]{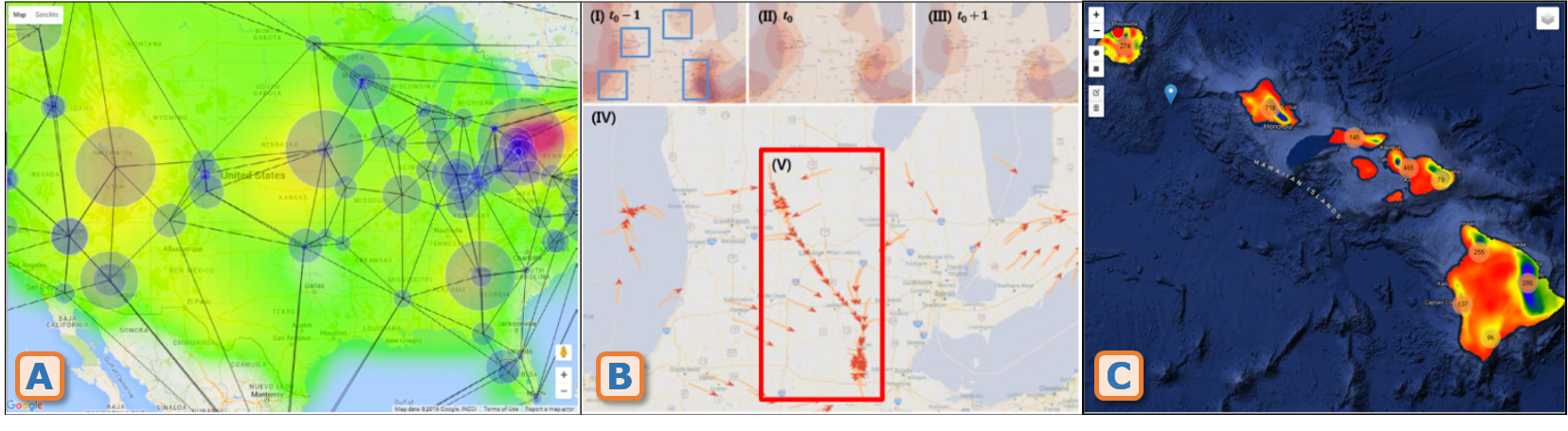}
    \caption{Examples of the color-coding technique leveraged in climate visualizations. (A) A geospatial heatmap and color-coded hotspots are merged with Delaunay triangulation to visualize the air pollution data. \cite{KALO2020169}. (B) Glyphs and color-coded hotspots are utilized to visualize data flow \cite{kim2018}. (C) HI-RAMA \cite{McLean2020} uses color schemes to present the rasterized station-wide rainfall data.}
    \label{fig:trajectory-vis}
\end{figure}

\subsubsection{Trajectory Visualization}
A trajectory visualization approach generally plots multiple temporal layers/glyphs into an integrated map where transparency and intensity derive the position of the object or event. This technique encodes the temporal dimension with different variations of glyph-based objects or colors as shown in Fig. \ref{fig:trajectory-vis}. To denote the spatial boundary, the trajectory technique is often merged with a map visualization \cite{kim2018, KALO2020169} or choropleth \cite{MacIejewski2010, Quinan2016}. Various transformations of this technique are leveraged to visualize the data flow, path ensembles, and trajectory in the spatiotemporal data.

Trajectory visualizations in climate science illustrate the temporal movement of climate attributes in geographical or abstract space. These visualizations can assist climate scientists to identify valuable patterns and trends from the spatiotemporal data. Various illustrations are employed to explore the underlying patterns in the moving object/event data \cite{10.1145/2093973.2094060}. Liu et al. \cite{liu2015visualizing} demonstrated trajectory visualizations for ensemble representations and position prediction of storm paths. They employed the ensemble path visualization to denote the hurricane direction and smoothly interpolated hotspots to denote the storm-strike position (Task requirement \textbf{R7}). Kim et al. \cite{kim2018} leveraged similar techniques to visually interpret the flow map in spatiotemporal data without trajectory information as shown in Fig. \ref{fig:trajectory-vis}B. Wang \cite{wang2014} published a software suite for meteorological data visualization, MeteoInfo, where the wind trajectory is demonstrated with ensemble path visualization over spatial contour from the station data.

Path ensemble simulations often leverage explicit encoding techniques to present high-dimensional and multivariate climate data. The spatial dimensions are usually encoded in a map, parallel coordinates, or hierarchical edge-bundling. In addition, the path ensemble information is commonly visualized using glyphs \cite{Sanyal2010}, hotspots \cite{liu2015visualizing}, flow maps \cite{kim2018}, volume rendering \cite{helbig2014}, etc. 






\subsubsection{Coordinated Multiple Views}
Coordinated multiple views (CMV) is an exploratory visualization technique that enables the user to explore the data using multiple visualizations integrated into a window \cite{roberts2007}. CMVs are employed to interactively visualize the complex high dimensional spatiotemporal data \cite{andrienko2007coordinated}. In spatiotemporal climate visualizations, CMV visualize the spatial and temporal aspect of the data using two different visualization techniques. A geospatial map is often used to visualize the spatial portion and a linear representation is used to visualize the temporal portion of the climate data \cite{mayr2018}. The goal is to allow the users to identify insights from complex spatiotemporal data by providing an integrated set of standard visualizations without cluttering the window. Visualizations in CMV are integrated through rich user interaction and view manipulation \cite{roberts2007}.

\begin{figure}[t]
    \centering
    \includegraphics[width=1.0\linewidth]{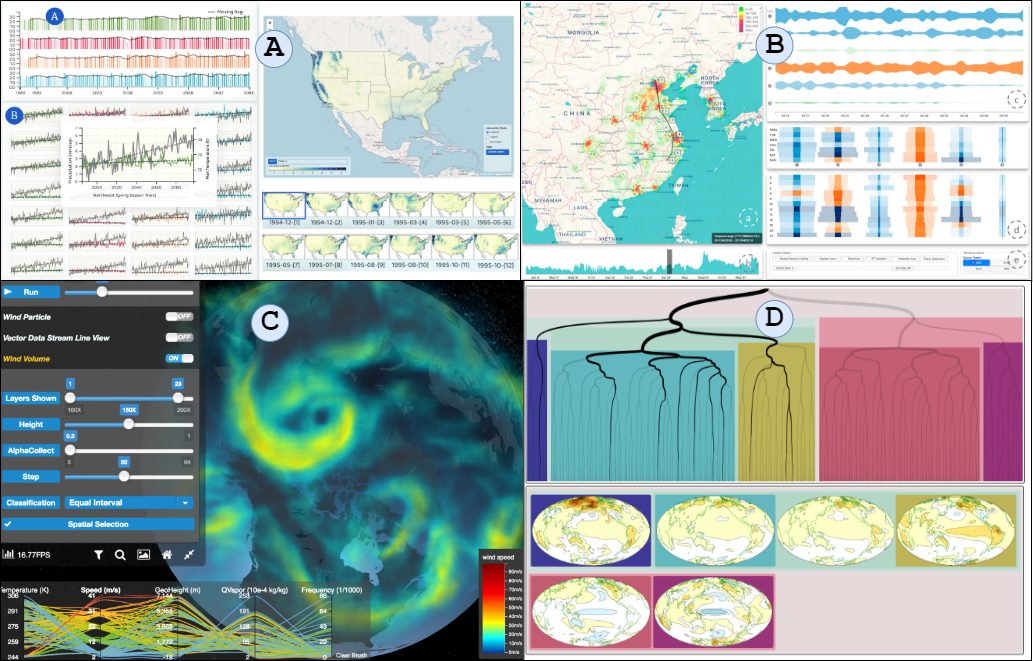} 
    \caption{Examples of CMV demonstrated in climate science VA to explore spatiotemporal data. (A) CMV leveraged to explore downscaled climate projections \cite{nayeem2021visual}. (B) DDLVis \cite{Lichenhui2021} facilitates real-time visual query of spatiotemporal data distribution. (C) Rule-based association in multivariate climate data is presented using a coordinated 3D globe and parallel coordinates \cite{wang2018}. (D) User-guided time-dependent hierarchical clustering to visualize the ensemble simulation \cite{kappe2019analysis}. }
    \label{fig:cmv-technique}
\end{figure}

Wang et al. \cite{wang2018} demonstrated CMV for conducting association rule-based multivariate analysis of spatiotemporal climate data. They employed a 3D volume rendering and parallel coordinates (Fig. \ref{fig:cmv-technique}C) to identify multivariate correlations among climate variables. Parallel coordinates are leveraged with geospatial spatial heatmap to compare multifaceted climate variables in SimilarityExplorer \cite{poco2014}. Li et al. \cite{Li2014} presented CMV in Vismate to visualize the station-based observation data on climate change. Moreover, Voila is a visual anomaly detection and monitoring system where Cao et al. \cite{Cao2018} presented a CMV compatible for visualizing the spatiotemporal climate variables. Li et al. \cite{Lichenhui2021} recently published a VA system, DDLVis, for the real-time visual queries of spatiotemporal data. They designed the interface with a geospatial map and area chart to navigate through the summarized spatiotemporal data. Based on their interaction with these visualizations the results are presented in a streaming view side-by-side, as shown in Fig. \ref{fig:cmv-technique}B.

CMV represents the spatiotemporal data using multiple visualization techniques that allow the user to split their focus into time and space dimensions \cite{mayr2018}. CMV often cases leverages the standard visualizations that fall under the visual literacy of the domain users. With rich user interactions and view manipulations, CMV overcomes the shortcomings of standard visualizations in extracting underlying knowledge from the data. However, depending on the visual encoding techniques, a hierarchical number of view manipulations, and user interactions, CMV can be overwhelming for the users. Several analytical reviews \cite{Convertino2003, andrienko2007coordinated, roberts2007} reported diversity in visual encoding and consistency in context switching are essential in visualizing spatiotemporal attributes in CMV.

\subsection{Interaction Methods}\label{sec:interaction-methods}
\begin{table*}
    \caption{Categories for the interaction methods in spatiotemporal climate visualizations. }
    \includegraphics[width=\linewidth]{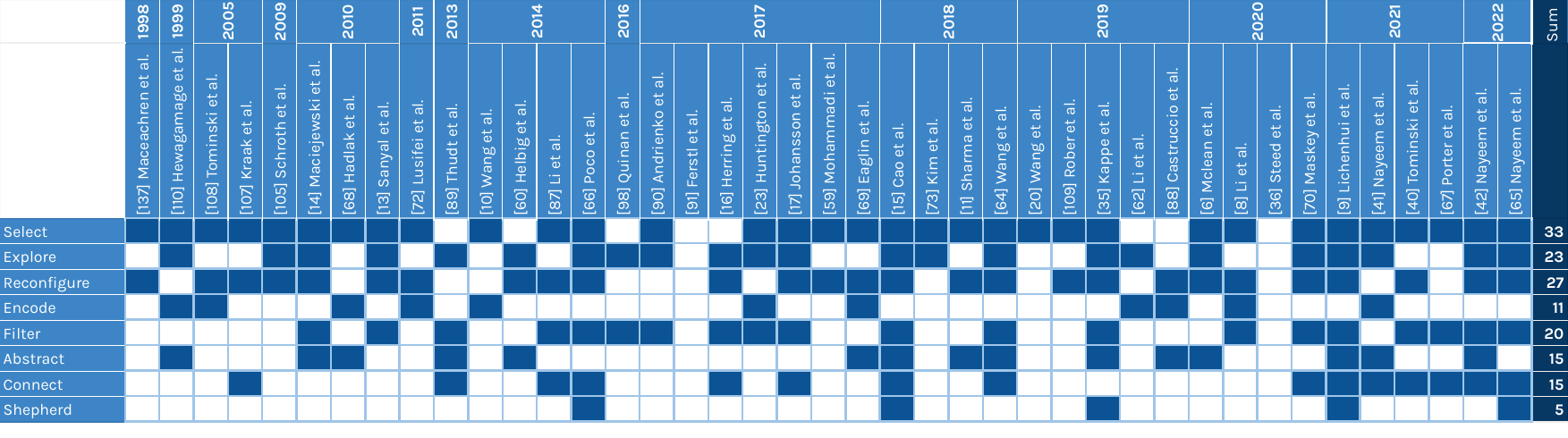}
    \label{tbl:cat-interaction-methods}
    \vspace{-5mm}
\end{table*}

User interaction in the visual analysis approaches is essential to facilitate the user's exploratory functionalities \cite{lu2017state}. What user interactions mean for visual analysis differs based on  systems, tasks, and user intentions \cite{dimara2019interaction}. In VA systems, user interactions affect the pipeline, transform raw and processed data, and alter the mapping and view \cite{dimara2019interaction}. In this section, we discuss different user interaction methods demonstrated in climate science VA. We categorize the interaction techniques based on the interaction methods taxonomy discussed in \cite{lu2017state, yi2007toward}. We also label the reviewed literature based on that taxonomy shown in Table \ref{tbl:cat-interaction-methods}. These user interactions are often correlated and overlapped while implementing the visualizations.


\textit{\textbf{Select}}
interaction enables users to mark or tag specific data based on their point of interest. 
This interaction method can be essential to remove outliers from the viewport.
In climate science VA, select is the most common interaction method. Panoply and NCCV \cite{Alder2015, 2018AGUFMIN21B35A} both render visualizations for climate projections where they offer selecting parameters for altering the time entity. NASA's Giovanni \cite{2007EOSTr8814A} provides interaction within the interface to select variables, time range and an analytical plot that essentially renders a static visualization. Select interaction is extensively leveraged in the CMVs as it allows the user to select an area of interest to update other visualizations accordingly \cite{Li2014, kappe2019analysis, nayeem2021visual}. McLean et al. \cite{McLean2020} employ select interaction on the map markers to display information about rainfall station and volume. Select interaction enables other interaction methods 
such as explore, connect, or abstract \cite{Sanyal2010, poco2014, eaglin2017space}. 

\textit{\textbf{Explore}} 
interaction provides users the control to include new items in the view, usually by superimposing them on the interactive view. Spatiotemporal climate variables are often multi-variate and multidimensional such that it is particularly challenging to display all the details in a single window. This interaction in such cases is very useful for the exploration of high-dimensional feature spaces.
Explore is a commonly used interaction method for integrated visualizations, especially for the map visualizations to support the identification of anomalies \cite{Cao2018}, uncertainties \cite{Sanyal2010}, or patterns \cite{kappe2019analysis} in the data. Quinan et al. \cite{Quinan2016} leveraged explore interactions to enable isocontour features on the ensemble map visualization. Johansson et al. \cite{Johansson2017} used this interaction to explore the risks for flooding and sea level rise in the context of climate scenarios for a selected location. 

\textit{\textbf{Reconfigure}} 
interaction enables the user to change the arrangement of visualizations by sorting or rearranging, useful for removing occlusion from the window. 
Reconfigure interaction is utilized to support time-varying large-scale multivariate analysis using parallel coordinates \cite{poco2014, li2020sovas}. Steed et al. \cite{Steed2020} demonstrated parallel coordinates visualization reconfigured with focus and context range selection for the temporal and numerical axes, as shown in Fig. \ref{fig:pcoords}. Wang et al. \cite{wang2018} leveraged reconfigure in 3D spatial volume rendering to demonstrate the rule-based multivariate analysis.

\textit{\textbf{Encode }} 
interaction fundamentally alters the visual representation by changing the number of dimensions, colors, and sizes in visualizations. Encode interaction can be particularly essential in multivariate analysis where the user can map the attributes to different colors and shapes.
Eaglin et al. \cite{eaglin2017space} employed encode interaction to allow the user to select a 2D temporal slice from the space-time cube. Li et al. \cite{Li2019} allow users to colorize the trajectories based on the variable and length of the trajectory. Moreover, interfaces employing parallel coordinates also demonstrate the use of encode interaction \cite{Steed2020} as they often change the color based on the selected range of multifaceted attributes.
    
\textit{\textbf{Filter}}
interaction helps the user to analyze data by applying conditions to the rendering. This interaction allows the user to specify a range or condition to visualize a subset of the data.
We reviewed a handful of VA systems employed in climate research that supports interaction to filter the sample space (Task requirement \textbf{R4}). DDLVis \cite{Lichenhui2021} employed filter interaction for the user to investigate an area of interest on the map visualization. Li et al. \cite{li2020sovas} allowed the user to write a textual query to visualize spatial heatmaps and temporal trends. Moreover, climate science produces complex and high dimensional data that often create occlusion in the view. The filter interaction is utilized in those scenarios to allow the user to focus on the data of interest \cite{MacIejewski2010, cli4030043, Cao2018, Maskey2020}. Sanyal et al.  \cite{Sanyal2010} allowed the user to filter the glyphs and ribbons presenting ensemble uncertainty in the geospatial visualizations. Climate Data Analysis Tool (CDAT) provides scope for user interaction to filter the data 
range that essentially reflects on the visualization \cite{potter2009visualization}.

\textit{\textbf{Abstract (or elaborate)}} 
interaction allows the user to view data from various levels of granularity. 
User interactions with the data points to see details in a tooltip or by zooming over the map are the common use cases of abstract interaction. 
Huntington et al. \cite{Huntington2017} employed abstract interaction in Climate Engine to allow the user to zoom on the time series and view values at the data points. Castruccio et al. \cite{castruccio2019visualizing} utilized abstract interaction in a portable virtual reality environment to explore climate variables across the globe and selected surfaces \cite{Sharma2018}.

\textit{\textbf{Connect}} 
interaction enables the user to find relationships between the views and highlight features that are similar or relevant. This interaction is applicable for both single-view and multiple views visualizations. In a single view, connect interaction highlights the related nodes or data points within the visualization. In multiple views, the user interacts with a record to effectively highlight all the related records across multiple views.
Climate science VA interface usually consists of multiple views \cite{Li2019, Maskey2020} as spatiotemporal data are often split into spatial and temporal dimensions to separate plots. VisAdapt \cite{Johansson2017} presents connect interaction where users interact with the geospatial contour to visualize the area detail on the other coordinated views. Wang et al. \cite{wang2018, cli4030043} demonstrate connect interaction where the spatial volume rendering updates based on the user interactions with the multivariate parallel coordinate plot.
    
\textit{\textbf{Shepherd}} 
is the interaction method that allows users to guide the modeling process. Such guidance can be direct or indirect. Direct Guiding enables the user to set or update different parameters for modeling whereas Indirect Guiding includes providing constraints and thresholds.
This interaction is not a common method. 
 Kappe et al. \cite{kappe2019analysis} demonstrated decadal climate prediction where they allowed the user to toggle the parameters to refine the ground truth in order to get the most accurate prediction. Cao et al. \cite{Cao2018} enable users to provide context-guided input for ranking anomalous patterns in streaming spatiotemporal data. Li et al. \cite{Lichenhui2021} allow users to guide real-time queries in spatiotemporal data distributions using density dictionary learning.

\section{Visual Analytics Systems}\label{sec:va-systems}
In this paper, we identify the task requirements of climate scientists for exploratory analyses with spatiotemporal data and review a comprehensive set of literature published in the visualization community addressing such requirements with a VA approach. VA includes the aspect of data preparation, pre-processing, analytical workflow, interactive visualization, and usability metrics of exploratory analysis. We review the VA approaches from the aspects of tools that are leveraged in the implementation process, system accessibility, and exploratory analysis workflow. 

\begin{figure}
\centering
\includegraphics[width=0.9\linewidth]{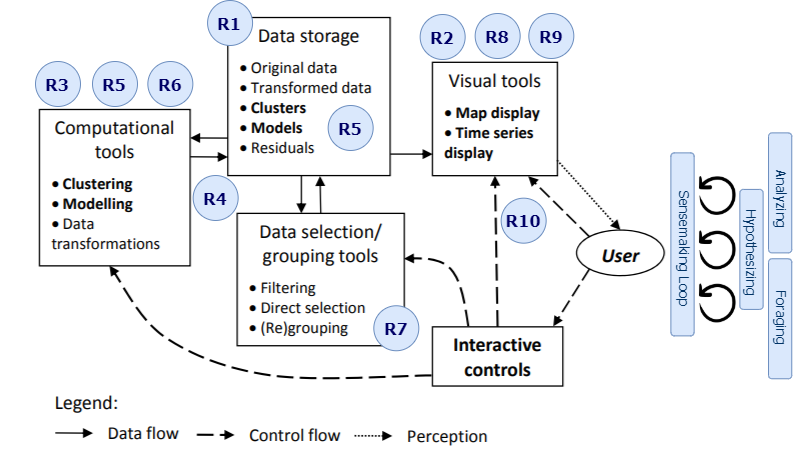}
    \caption{Spatiotemporal VA workflow for exploratory analysis \cite{andrienko2013visual}. We contextualized the workflow for climate science by labeling it with our identified task requirements for exploratory spatiotemporal VA.}
    \label{fig:exploratory-model}
\end{figure}

Cashman et al. \cite{cashman2019user} presented a user-based VA workflow for exploratory analysis that enables the users to discover underlying knowledge from the given dataset. The workflow consists of several fundamental steps such as data interaction, problem exploration, problem specification generation, and model exploration. Cui \cite{Cui2019} provided a comprehensive analytical review of the VA systems and summarizes the analytics process in 6 steps - data pre-processing, analysis method, visualization, knowledge generation, interactive hypothesis building, and visual reflection of user perception. Moreover, Nayeem et al. \cite{nayeem2021visual} demonstrated a VA framework for distributed analysis where users engage in a sensemaking loop by iteratively hypothesizing, analyzing, and foraging through data exploration, analyses, and interactive visualizations. Sacha et al. \cite{sacha2014} illustrated a knowledge generation model for VA that represents both computer and human-centric approaches through exploration, verification, and sensemaking loop. Andrienko et al. \cite{andrienko2013visual} described a VA framework for spatiotemporal analysis and modeling. In addition to the conventional exploratory VA workflow, this research addressed the user's perception of data and control flow from a spatiotemporal analysis and visualization perspective. Based on our extensive literature review, we believe Andrienko's VA illustration \cite{andrienko2013visual} more accurately outlines the VA workflow for spatiotemporal exploratory analysis in climate science. We have labeled their illustration of the main VA components with our identified task requirements (Section \ref{sec:task-requirements}) as depicted in Figure \ref{fig:exploratory-model}.

While reviewing the implementation process, we found that VA systems often leverage other data analytics tools to run exploratory analyses. These tools mostly enable climate scientists to run analytical scripts and visualize the result with basic visualizations. It is popular among climate scientists to leverage tools such as MATLAB \cite{hanselman1996mastering}, GrADS \cite{Berman2001}, or NCAR Command Language (NCL) \cite{NCAR2019}. Moreover, exploring spatiotemporal data frequently leverages the capabilities of distributed computing \cite{Alder2015, kim2018, Cao2018, Lee2018regional, nayeem2021visual}, to overcome the complex and time-consuming analysis. Several systems are widely used by analysts and research communities that provide shared computing resources to run analyses in the spatiotemporal data. The Ultra-scale Visualization Climate Data Analysis Tool (UV-CDAT; \cite{santos2013uv}) is one of the oldest systems among those reviewed in this report. However, UV-CDAT includes statistical analysis tools to support the specific needs of the climate science community (Task requirement \textbf{R5}), such as those used in the Assessment Reports from the Intergovernmental Panel on Climate Change (IPCC). DataONE is another system for the earth and environmental science data repositories to facilitate easier access, search, and discovery \cite{sandusky2016computational, cohn2012dataone}. The Open Science Grid serves the scientific research community with its large-scale distributed computing resources, on more than 120 institutes \cite{pordes2007open}. In addition, SciServer is a cyberinfrastructure that provides access to a suite of tools and services. SciServer infrastructure includes storage, query, and processing for big data analyses from different domains with diverse data formats and structure \cite{medvedev2016sciserver}. We have not found implementations of DataONE, Open Science Grid, and SciServer in climate science VA. Virtual Information Fabric Infrastructure (VIFI) is another distributed data analysis system that enables computation without moving the massive data \cite{8397642}. Within VIFI, the analytic scripts move over to the data hosts, produce the resulting data, and return to the user site. The potential of VIFI in analyses of complex spatiotemporal data has been demonstrated in \cite{nayeem2021visual}. These analysis tools usually lack interactive exploratory visualizations and are unable to satisfy climate scientists' custom visualization requirements. In contrast, VA systems enable climate scientists to perform their analysis tasks and explore complex spatiotemporal data with the support of an interactive interface. 

We label the climate science VA systems in context to exploratory spatiotemporal analysis into two categories - general purpose and case-specific (Table \ref{tbl:va-applications}).
General-purpose VA systems support analysis for numerous use cases and datasets. These systems generally demand the dataset be organized according to a pre-defined structure. Climate Engine \cite{Huntington2017}, SOVAS \cite{li2020sovas}, and WebGlobe \cite{Sharma2018} are some of the leading examples of general-purpose VA systems we reviewed in this paper. However, these systems often fall short to satisfy specific analytics or visualization requirements for case-specific VA systems. Climate scientists work with heterogeneous data that frequently demand custom features to analyze and explore the underlying trends and patterns. Most special purpose systems have been developed by domain experts with their specific requirements \cite{nocke2008climate}. In our review, we also observed climate scientists collaborating with visualization researchers to develop VA systems that support case-specific exploratory analysis of climate research (Task requirement \textbf{R5, R9}). Noodles \cite{Sanyal2010}, VisAdapt \cite{Johansson2017}, DDLVis \cite{Lichenhui2021} demonstrate examples of case-specific VA systems. 

The modern web platform enables visualization researchers to develop large-scale web-based VA systems that efficiently integrate data management, transformation, hypothesis testing, and knowledge discovery with interactive visualizations \cite{kalinin2017}. In contrast, native VA systems execute the analytics and render visualizations within a common computational infrastructure without the movement of the data. While the native approach is built on a relatively less complicated application design and ensure real-time access to the data, it is often challenging in the context of complex spatiotemporal analysis to move massive data and leverage high-performance computing resource. On the other hand, web-based systems generally utilize cloud-based high-performance clusters to execute the analytics. Climate scientists also exploit the advantages of cloud-based shared computing resources to run their analysis \cite{Nemani2015, 8397642, Lee2019}. Hence, web-based systems obtain the potential to support a modular architecture and protocol for simultaneous accessibility to analysis infrastructure without the movement of a massive volume of data (Task requirements \textbf{R1, R2, R4, R8}). In our survey, we also coded the current VA approaches in terms of native and web-based accessibility and presented them in Table \ref{tbl:va-applications}. 

We identify the tools and libraries that are commonly used in the development of visualizations and analytics interfaces. We found several examples where native applications are built using Visual C++ and visualizations are developed utilizing OpenGL \cite{hewagamage1999interactive, Quinan2016}. In recent years, the trend progressed more towards web-based VA systems as visualization researchers leverage JavaScript-based libraries such as AngularJS, and ReactJS to develop an interactive interface \cite{Johansson2017, McLean2020}. Standard 2D/3D Visualizations are developed using WebGL, D3.js, ThreeJS in several analytics interfaces \cite{kim2018, Sharma2018, li2020sovas}. It is evident from our discussion in section \ref{sec:visualization-techniques} that map visualizations hold a crucial aspect in exploratory spatiotemporal visual analysis. Google Earth, ArcGIS, and LeafletJS are commonly leveraged to develop map visualizations \cite{Huntington2017, Alder2015, nayeem2021visual} for the VA interfaces.



\begin{table*}
    \caption{Categories for evaluation methods. Use cases are found as the most preferred method 
    in climate science research.}
    \includegraphics[width=\linewidth]{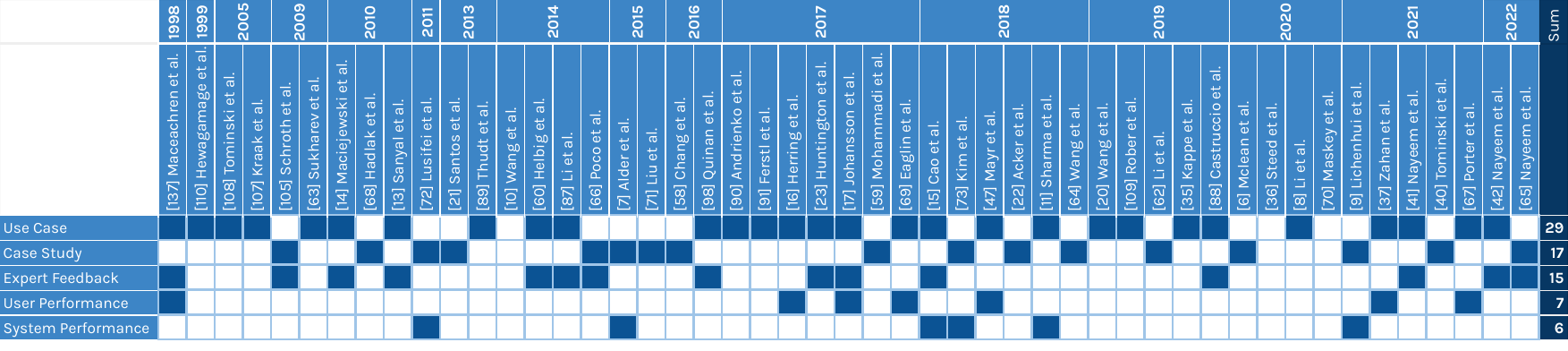}
    \label{tbl:cat-evaluation-methods}
    \vspace{-5mm}
\end{table*}

\section{Evaluation Methods}\label{sec:evaluation-methods}
Evaluation of the visualization techniques and analytics systems is an essential task as it can validate the contribution of the proposed tool in climate science research. Johansson et al. \cite{johansson2010evaluating} reflected on the approaches for evaluating climate visualization in order to support advanced interactive visual analysis. They combined three aspects of climate visualizations - analysis, communication of research results, and decision support to evaluate their utility in understanding the complex data. These aspects provide actionable insights to consider more qualitative and quantitative evaluation metrics for scientific evaluation. Kristensson et al. \cite{Kristensson2009} discussed the evaluation approach of spatiotemporal representations taking user performance, task completion time, and user experience into account. In terms of quantitative evaluation, Mayr et al. \cite{mayr2018} acknowledged that the conventional method of evaluating spatiotemporal visualizations with time and error indicators often falls short. Comparative evaluations, usability with multivariate information, and participants' visual literacy are all crucial for a justifiable evaluation \cite{nocke2007visual}. Moreover, Elmqvist et al. \cite{elmqvist2015patterns} discussed several patterns for visualization evaluation including exploration, control, generalization, validation, and presentation. The anatomy of each pattern consists of five basic components including evaluation name, problem context, proposed solution, consequences, and illustrative examples. 

We take the evaluation patterns \cite{elmqvist2015patterns} into account to categorize the evaluation methods of the reviewed VA approaches based on expert feedback, case study, use cases, performance, and the user's ability to comprehend without extensive training (\emph{do-it-yourself}). The case study describes the scenario where the system is demonstrated with the help of domain experts conducting specific analysis tasks. In contrast, use cases include a number of hypothetical usage scenarios where the proposed system might be leveraged in performing analysis tasks. Table \ref{tbl:cat-evaluation-methods} presents an overview of the evaluation methods utilized in the VA interfaces related to spatiotemporal climate research.

MacEachren et al. \cite{729563} conducted a quantitative evaluation to assess the temporally varying geographic visualization. Participants are provided with 3 sets of tasks to analyze spatiotemporal patterns. A timeline map is presented for each participant that marks the illustration of every interactive feature utilized to perform the analysis tasks. Herring et al. \cite{Herring2017} provided an online portal to conduct a quantitative performance evaluation of a geospatial interface for US climate data with 55 participants. The proposed interface was evaluated based on each participant's time of assessment and accuracy which entails the perceived reality of climate change. Mayr et al. \cite{mayr2018} conducted a comparative quantitative analysis of the 4 different integrated visualizations in performing spatiotemporal analysis. Inferential statistical techniques such as t-tests, and analysis of variance (ANOVA) are widely used in the comparative assessment of the performance metrics \cite{Herring2017, eaglin2017space, Zahan2021}. 

As we discussed in Section \ref{sec:va-systems}, the domain expert involvement in the design and development process \cite{castruccio2019visualizing} of case-specific VA systems is valuable. Their feedback is influential in the evaluation process as well. Sanyal et al. presented a visualization tool, Noodle \cite{Sanyal2010}, for exploring ensemble uncertainty in numerical weather models. Two meteorologists were included in the evaluation process where they used the tool to perform ensemble analysis and provided qualitative feedback. Several research works have taken similar steps to evaluate proposed systems \cite{Quinan2016, Steed2020}. Moreover, domain experts are also invited to participate in the evaluation and present pre-defined use cases to validate the usability and novelty of the VA interfaces \cite{Cao2018, Huntington2017}.

In our review, we found that use cases are the most popular method for evaluation, followed by case studies. In contrast, we did not find any examples where a user's ability to comprehend the analytics interface on their own (\emph{do-it-yourself} \cite{elmqvist2015patterns}) is considered as an evaluation metric. Johansson et al. conducted the evaluation for VisAdapt \cite{Johansson2017} using an iterative process including both climate scientists and end-users from diverse backgrounds. According to their report, familiarity with the interface is essential for the users to leverage the VA system to explore the climate change impact and possible adaptive action.

\begin{figure}[t]
    \centering
    \includegraphics[width=0.95\linewidth]{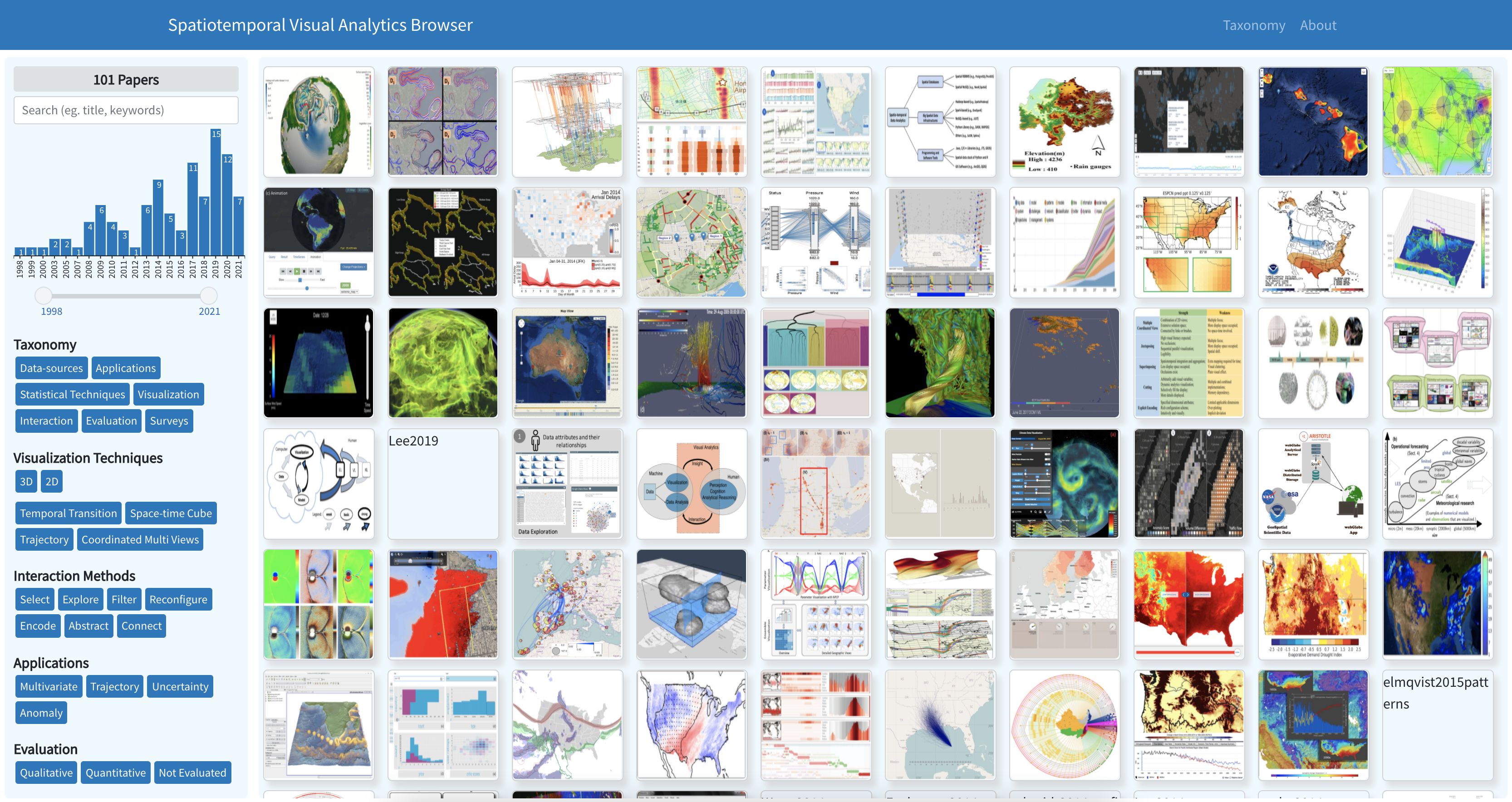}
    \caption{A web interface to browse the literature we surveyed in this report related to VA approaches for climate science.}
    \label{fig:paper-browser}
\end{figure}


\section{Summary \& Discussion}\label{sec:discussion}
We surveyed the progress in exploratory spatiotemporal visualization and VA approaches in climate science to date and intend to outline the potential future direction to address the present research gaps.
To establish the scope of our survey, we first reviewed the previous surveys related to our topic \cite{yang2019research, CHEN2019129, andrienko2020spatio, alam2021survey}. These compiled surveys and analytical reviews show the overall advancements in spatiotemporal visualization and VA. However, they lack context for advanced data analytics for climate scientists. The existing surveys specific to the climate science visualizations \cite{Tominski2011cr, Rautenhaus2018} are quite outdated as they include papers published on or before 2018. Since then, significant advancements have been made in exploratory spatiotemporal VA in climate science. Moreover, surveys related to climate science reported their analytical review mainly in the context of data analysis methods and visualization techniques. In contrast, we report the exploratory spatiotemporal analysis in climate science more from the VA perspective. By doing so, we can identify the key aspects of spatiotemporal VA to present a taxonomy that includes the data sources, analysis tasks, statistical techniques, visualization techniques, interaction methods, and evaluation methods, all of which facilitate exploratory data analysis in climate science.

We implemented an interactive web interface to present the papers that are included in the current report. In our interface, we placed each paper as a thumbnail with the figure that represents the paper best. The user can hover over the thumbnail to see the paper's title. To see more details about the paper such as authors, abstract, keywords, publication venue, and year, the user can click on the thumbnail to open a popup window. We tagged the papers based on the categorizations we discussed in this paper. The interface leverages the tags to allow the user to filter the paper based on their selection on the left side panel. Figure \ref{fig:paper-browser} shows an illustration of our web interface
\footnote{The interface for spatiotemporal climate VA browser: \url{http://vizus.cs.usu.edu/app/climatescience}
}

\subsection{Knowledge Gaps \& Future Scope}
While we reflect on the current trends in exploratory spatiotemporal VA research related to climate science, we identified existing gaps and potential future research directions to address these knowledge gaps and the ongoing challenges of understanding spatiotemporal climate data. We summarize some of the key aspects below:

\textbf{Hierarchical clustering for visualizing spatiotemporal granularities.} In our data type classification, we found a majority of climate data falls into a multidimensional type due to the high dimensionality, multi-variability, and multi-resolution factors in climate data. In contrast, tree-type datasets are not much utilized in climate VA. However, we understand interpreting spatiotemporal data to tree data type holds enormous potential to observe patterns and identify correlations among climate variables through several spatial (e.g., region, state, and county) and temporal granularities (e.g., annual, seasonal, and monthly) (Task requirement \textbf{R7}). Consequently, hierarchical clustering \cite{kappe2019analysis} (depicted in Fig. \ref{fig:cmv-technique}D) and treemap \cite{Schreck2006} visualization techniques are not extensively utilized in spatiotemporal climate data exploration and have the potential to be a new research
direction for climate science visualizations.

\textbf{Growing trend of integrated visualizations in spatiotemporal exploration.} In recent years, integrated visualization techniques have become widely utilized as tools to visualize spatiotemporal climate data. Integrated visualizations are essential for the exploratory analysis of climate variables because of their usability and existing and potential future new interactive features. Visualizations that enable scientists to perceive salient data features quickly and efficiently are key. For example, distributing the spatial and temporal dimensions across multiple visual components allows users to inspect the data by focusing on individual dimensions. Consequently, such a capability not only helps scientists to identify underlying trends and patterns from the data but also helps them to maximize cognitive efficiency. Table \ref{tbl:cat-vis-techniques} suggests CMV has become a trending approach for the interactive exploration of large-scale spatiotemporal climate data. However, significant challenges remain in terms of choosing relevant visualization techniques, display formation, and interactive features for exploratory visual analysis. Roberts \cite{roberts2007} reflected on the view generation, interaction, and manipulation aspect to achieve state-of-the-art in CMV.


\textbf{Underexplored immersive visual environments.} Climate scientists often use point cloud data from observations and numerical models which can be analyzed in a fully immersive environment \cite{marriott2018immersive} applying Virtual and Augmented Reality technologies. Such an immersive analytic environment would provide better spatial perception to climate scientists for exploratory spatial analysis with other climate variables and a collaborative environment to discuss and share their findings with other scientists and researchers.

\textbf{Interactive shepherding of exploratory models.} Table \ref{tbl:cat-interaction-methods} suggests that shepherd interactions are underutilized in climate science VA. This entails a shortcoming of these systems in facilitating user-driven exploratory analysis. Systems such as UV-CDAT \cite{santos2013uv} and MeteoInfo \cite{wang2019open} comprise active participation in building the exploratory analysis model. However, these systems are heavily code dependent and often lack interactive features (Task requirement \textbf{R10}). This unfolds a research opportunity to enable interactive user-driven exploratory analysis from the VA interface (Task requirement \textbf{R3}).

\textbf{Case-specific approaches over general-purpose VA.} We grouped the spatiotemporal VA systems for climate science into general-purpose and case-specific systems. While the general purpose systems \cite{Sellis2011, santos2013uv, Huntington2017} are well-commended for diverse usability and rich visualization library, these systems often lack the domain or use-case-specific custom visualizations and user interactions to support analysis tasks or demand high levels of programming expertise from the user. In addition, we reviewed VA systems developed to serve specific analysis tasks or data sources \cite{McLean2020, KALO2020169, Lichenhui2021, dcpviz2022}. In contrast to the general purpose systems, these systems are more oriented to employ a use-case-specific analytics pipeline, visualization techniques, and user interaction. Climate scientists and visualization researchers usually engage in a joint collaboration to develop case-specific systems to satisfy the analysis requirements. Table \ref{tbl:va-applications} indicates a significant number of VA systems developed for climate science focus on specific use cases (Task requirement \textbf{R9}). 

\textbf{Surge of web-based VA systems.} As climate scientists work with diversified datasets to understand climate change and its social impact and derive effective decision-making for adaptation, they often obtain data from disparate sources to perform distributed decision support and sensemaking analysis \cite{Cao2021}. To conduct such analyses, they commonly transfer the data from distributed sources to their local workstations, apply data processing and transformations on the workstations, then generate visual illustrations. To mediate this inefficient workflow, a web-based VA approach can address these challenges of accessing disparate remote data sources, performing case-specific analyses on high-performance clusters, and providing an interactive interface for exploratory spatiotemporal visualizations \cite{Porter2021, dcpviz2022} (\textbf{R2, R8, R10}). With advancements in modern web technology, VA systems are widely built on the web-based platform in recent years as shown in Table \ref{tbl:va-applications}.

\textbf{Low pervasion among climate scientists.} A comprehensive review of several analytical reports and surveys \cite{Tominski2011cr, afzal2019state}, as well as, our discussion with the atmospheric and environmental scientists identify that most climate scientists do not leverage state-of-the-art VA solutions to conduct their exploratory analyses. This gap between VA capabilities and climate science advancement must be bridged. Traditional approaches that incidentally rely on visualizations are becoming increasingly insufficient to match the scope of the climate science community's scientific challenges. Much of the information contained in the ever-growing observational and modeling datasets remains untapped. For example, two-dimensional plots, that are central to publications and anchor scientific paper development are highly reductive. Training of climate scientists and collaboration with VA domain experts will help bridge these persistent gaps.  

\textbf{Unexploited potential of progressive visualization.} 
There are growing numbers of observational platforms with increased data rates, growing numbers of climate models that exhibit increasing complexity, and a growing sophistication of experiments undertaken with climate models \cite{Eyring2016CMIP6}. This rapid growth of data also challenges VA systems to support on-demand exploratory capabilities over large volumes of data. Conventional methods in VA systems often leverage background computation based on user queries \cite{Ma2009}. They, then, display the computed result at once upon completion. Zgraggen et al. \cite{7563865} labeled such a method as a `blocking approach' where the user's analytical visions are blocked until the computation is fully completed. To address this challenge, climate research can benefit from the potential for progressive visualizations where data are incrementally processed in smaller chunks as the analytics systems interactively update the interface, working from an approximation that is refined over time \cite{7563865, gmd-12-233-2019} (Task requirements \textbf{R1, R4}). As the computations take place in chunks, shepherd interaction \cite{lu2017state} can be utilized to build the exploratory analysis model. In our survey, we have not found any progressive VA systems to explore spatiotemporal climate data. This indicates another opportunity for visualization researchers to develop progressive systems in collaboration with climate scientists.




\section{Conclusion}
Our paper presents a survey on the state-of-art exploratory spatiotemporal VA approaches for climate science. We identified the task requirements, reviewed a comprehensive set of literature based on the taxonomies, and presented the advancements in the exploratory analysis of climate science from the VA perspective. We also uncover potential research opportunities and the need for collaboration among climate scientists and visualization researchers to bridge gaps between the use of visualizations in traditional climate science and the capabilities of visualization researchers. The collaboration will likely become ever more urgent in the face of growing data volumes where the lack of efficient means to extract information from datasets will result in that information being unused. Consequently, we believe that our survey provides useful guidance to enable bridge-building between VA researchers and climate scientists to produce systems that enable deeper explorations of 
climate change data that will be increasingly necessary to advance the scientific understanding of these domains.   

\section*{Acknowledgements}
This paper was supported by the US National Science Foundation (NSF) Data Infrastructure Building Blocks (DIBBs) Program (Award \#1640818).

\ifCLASSOPTIONcaptionsoff
  \newpage
\fi



\bibliographystyle{IEEEtran}
\bibliography{main}


\begin{IEEEbiographynophoto}
{Abdullah-Al-Raihan Nayeem}
Abdullah-Al-Raihan Nayeem is a Ph.D. candidate in Computer Science at the University of North Carolina at Charlotte. His research interests are visual analytics systems, geo-spatiotemporal visualization, and distributed data analysis. 

\end{IEEEbiographynophoto}

\begin{IEEEbiographynophoto}
{Dongyun Han}
Dongyun Han is a Ph.D. student in the department of Computer Science at Utah State University. His research interests include visual analytics, data visualization, and human-computer interaction. 
\end{IEEEbiographynophoto}

\begin{IEEEbiographynophoto}
{Huikyo Lee}
Huikyo Lee joined Jet Propulsion Laboratory as a data scientist. 
He has a Ph.D. in Atmospheric Sciences from the University of Illinois at Urbana-Champaign.
He has 
had extensive experience with stratospheric dynamics, neural network modeling for satellite remote sensing, atmospheric chemistry, and air quality modeling. 
\end{IEEEbiographynophoto}
\begin{IEEEbiographynophoto}
{Donghoon Kim}
DongHoon Kim 
is a PhD student in the Computer Science Department at Utah State University. His research interests are human perception in virtual reality and visualization.
\end{IEEEbiographynophoto}

\begin{IEEEbiographynophoto}
{Daniel Feldman}
Daniel Feldman is a scientist in the Earth and Environmental Sciences Area at the Lawrence Berkeley National Laboratory.  He has a Ph.D. in Environmental Science and Engineering from Caltech, and a B.S. in Environmental Engineering Science from MIT. He research focuses on the intersection between Earth System Modeling and observations. 
\end{IEEEbiographynophoto}

\begin{IEEEbiographynophoto}
{William J. Tolone}
William J. Tolone is a professor and associate dean of the College of Computing and Informatics at the University of North Carolina at Charlotte. He received his Ph.D. in Computer Science from the University of Illinois at Urbana-Champaign. His research expertise is in integrated modeling and simulation, critical infrastructure analytics, visual and data analytics, and collaborative systems. 
\end{IEEEbiographynophoto}

\begin{IEEEbiographynophoto}
{Daniel Crichton}
Daniel Crichton is a program manager, principal investigator, and principal computer scientist at NASA's Jet Propulsion Laboratory, which he joined in 1995. He is JPL’s leader of the Center for Data Science and Technology, a joint center formed with Caltech, focusing on the research, development and implementation of data science and data intensive system technologies for science and missions.  
\end{IEEEbiographynophoto}

\begin{IEEEbiographynophoto}
{Isaac Cho}
Isaac Cho is an assistant professor in the Computer Science department at Utah State University and an adjunct professor in the Computer Science department at the University of North Carolina at Charlotte. His main research interests are interactive visual analytics, data visualization, and human computer interactions. 
\end{IEEEbiographynophoto}








\end{document}